\documentclass[10pt,a4paper]{article}
\usepackage[utf8]{inputenc}
\usepackage[T1]{fontenc}
\usepackage{hyperref}
\usepackage[nottoc]{tocbibind} 
\usepackage{fancyhdr}
\usepackage{graphicx}
\usepackage{color}
\usepackage{bm} 
\usepackage{amsmath}
\usepackage{amssymb} 
\usepackage{amsthm}  
\usepackage{mathrsfs} 
\usepackage{ifthen}
\usepackage{tikz}
\usepackage[framemethod=TikZ]{mdframed} 

\newtheorem{definition}{Definition}

\newtheorem{theorem}{Theorem}
\newtheorem{proposition}{Proposition}

\newcommand{\w}[1]{\pmb{#1}}  

\newcommand{\B}[1]{\overline{#1}}
\newcommand{\ts}{\otimes}
\newcommand{\Tld}[1]{\widetilde{#1}}
\newcommand{\Div}{\w \nabla \cdot}

\newcommand{\la}{\langle}
\newcommand{\ra}{\rangle}

\newcommand{\wnab}{\w{\nabla}}

\newcommand{\dd}{\mathbf{d}}

\newcommand{\defin}[1]{\textbf{\itshape #1}}
\newcommand{\be}{\begin{equation}}
\newcommand{\ee}{\end{equation}}
\newcommand{\bea}{\begin{eqnarray}}
\newcommand{\eea}{\end{eqnarray}}
\newcommand{\bsea}{\begin{subeqnarray}}
\newcommand{\esea}{\end{subeqnarray}}

\newcommand{\Liesymbol}{\mathcal{L}}

\newcommand{\RN}[1]{\textup{\uppercase\expandafter{\romannumeral#1}}}

\newcommand{\R}{\mathbb{R}}
\newcommand{\Z}{\mathbb{Z}}

\newcommand{\M}{\mathscr{M}}
\newcommand{\D}{\mathscr{D}}

\newcommand{\g}{\w{g}}

\newcommand{\T}{\mathrm{T}}

\newcommand{\spose}[1]{\hbox to 0pt{#1\hss}}
\newcommand{\lta}{\mathrel{\spose{\lower 3pt\hbox{$\mathchar"218$}}
     \raise 2.0pt\hbox{$\mathchar"13C$}}}
\newcommand{\gta}{\mathrel{\spose{\lower 3pt\hbox{$\mathchar"218$}}
     \raise 2.0pt\hbox{$\mathchar"13E$}}}

\newcommand{\N}{\mathscr{N}}
\newcommand{\C}{\mathscr{C}}
\newcommand{\U}{\mathscr{U}}
\newcommand{\F}{\mathscr{F}}
\newcommand{\V}{\mathscr{V}}
\newcommand{\W}{\mathscr{W}}

\newcommand{\Hor}{\mathscr{H}}

{\begin{svgraybox}\begin{example}}%
{\end{example}\end{svgraybox}}

 {\small\begin{remark}}%
 {\end{remark}\normalsize}

\newmdenv[backgroundcolor=gray!20!white,roundcorner=5pt,hidealllines=true]{greybox}

\begin{document}

\title{A new approach to Kaluza-Klein Theory.}
\author{Stephane Collion and Michel Vaugon.\\ \sl \small Institut de Math{\'e}matiques, Universit{\'e} Paris VI, Equipe
  G{\'e}om{\'e}trie et Dynamique, 
    \\ \sl \small email: vaugon.michel@wanadoo.fr, stephane.collion@wanadoo.fr}
\date{August 2017, August 2022}
\maketitle

\begin{abstract}

We propose in this paper a new approach to the Kaluza-Klein idea of a five dimensional space-time unifying gravitation and electromagnetism, and extension to higher-dimensional space-time. By considering a natural geometric definition of a matter fluid and  abandoning the usual requirement of a Ricci-flat five dimensional space-time, we show that a unified geometrical frame can be set for gravitation and electromagnetism, giving, by projection on the classical 4-dimensional space-time, the known Einstein-Maxwell-Lorentz equations for charged fluids. Thus, although not introducing new physics, we get a very aesthetic presentation of classical physics in the spirit of general relativity. The usual physical concepts, such as mass, energy, charge, trajectory, Maxwell-Lorentz law, are shown to be only various aspects of the geometry, for example curvature, of space-time considered as a Lorentzian manifold; that is no physical objects are introduced in space-time, no laws are given, everything is only geometry. 

We then extend these ideas to more than 5 dimensions, by considering spacetime as a generalization of a  $(S^1\times W)$-fiber bundle, that we named \emph{multi-fibers bundle}, where $S^1$ is the circle and $W$ a compact manifold. We will use this geometric structure as a possible way to model or encode deviations from standard 4-dimensional General Relativity, or "dark" effects such as dark matter or energy.

\footnotesize{This paper is a rewriting and improvement of a previous joint work, [20],  with B. Vaugon, M. Dellinger, Z. Faget.}
\footnote{Version 02 08 2022. AMS subject classification: 83C22, 83E05, 83E15}
\end{abstract}

\section{Introduction}
The axiomatic of General Relativity is beautiful and very simple, so long as one does not attempt to fully introduce electromagnetism in it. Indeed, when considering the so-called perfect fluids, describing electrically neutral matter, all the physics is described by a geometrical setting, a Lorentzian manifold, and a natural "energy-momentum" tensor, which is set to be equal to the Einstein curvature of the manifold. The equations of motion as well as the conservation laws are then given by a purely geometrical theorem: the Bianchi identity. And that's it. One get all the fundamental aspects of gravity: movement of planets (through Scharzschild metric), Big Bang, Light deviation, black holes. (We will recall this in the second section). Gravitational free-fall is just geodesic movement.

However, when one wants to include the other part of classical physics, electromagnetism, for which special relativity was "invented", new object and laws are to be introduced : a closed differential 2-form $F$ must be given on the space-time manifold, the Maxwell laws are to be postulated, and a not-so-natural energy momentum tensor must be given, which do not derive easily from the 2-form.

The goal of this paper is to propose a geometrical setting, based on the now classical but still brillant idea of Kaluza and Klein, where gravitation and electromagnetism are united, where the classical physical concepts, such as mass (Baryon number), energy, charge, are just characteristics of the geometry of the space-time-manifold, and where classical physical laws of movement as well as Maxwell-Lorentz laws are given by purely geometrical facts (theorems), mainly the Bianchi identity. Moreover, the fundamental equivalence principle, stating that free particles follow geodesics of space-time, will be generalized to the case of charged particles in a gravitational and electromagnetic field, that will be shown to follow geodesics of a 5-dimensional space-time.

The main ideas on which this work is based can be summarized in the following way :

A. Space-time is a pseudo-Riemannian manifold. It can be considered as an approximate model of a more complicated sub-structure of nature, used to introduce the extremely powerful tools of differential calculus.

B. Physical objects and data we observe are just peculiar aspects of the geometry of some domain of the space-time manifold ; for example characteristics of its curvature.

C. Laws of physics, such as conservation laws or equations of motion, are just geometrical theorems, such as the Bianchi identity, applied to the above observed domains of the space-time manifold.

We will apply this frame in three steps. First to classical general relativity to expose our ideas in a known context. Secondly to a five-dimensional space-time in the spirit of Kaluza-Klein, with the simple case of dust fluid, to unite gravitation and electromagnetic in a common geometrical setting. Thirdly, we will apply this frame to a generalization of the notion of fiber bundle, that we called \defin{multi-fibers bundle}, and show that 
this geometric structure is a possible way to model or encode deviations from standard 4-dimensional General Relativity, or "dark" effects such as dark matter or energy.
\\

\footnotesize{ \emph{In all the paper,  $(\M,\g)$ is  a semi-Riemannian manifold with metric $\g$, $\wnab$ is its Levi-Civita connection, $\w R$ and $\w{Ric}$ are the Riemann and Ricci curvatures. $\T_p\M$ is the tangent space of $\M$ at $p$ ; $T_pf$ is the tangent map at $p$ of a function $f$. We sometimes note the scalar product $\g(X,Y):=\la X,Y \ra$. We note $\Div \w T$ the divergence of a tensor $\w T$. In all the paper, using musical isomorphisms, we identify without any comments a $(0,2)$-tensor  $\w T$ with the $(2,0)$-tensor $\w T^{\sharp \sharp}$. They represent the same physical object. In particular, the Einstein curvature $\w G=\w{Ric}-1/2\w S.\g$ will often be consider as a $(2,0)$-tensor, $\w G=G^{ij}$. We also consider, for a 2-tensor $\w T$, the divergence $\Div \w T$ as a vector, that is, we identify the 1-form $\Div \w T$ and $(\Div \w T)^\sharp$. At last,  we shall note $^e \w T$ the endomorphism field $\g$-associated to $\w T$ : 
 $\forall (\w{u},\w{v}) \in \T_p(\M)\times\T_p(\M)\,$ :  $\, \g ( \w{u} , ^e\w{T}(\w{v}))  =\w{T}(\w{u},\w{v})$.}}
 \normalsize
 
\section{General Relativity.}
Let $(\M,\g)$ be a spacetime, that is, a time-oriented Lorentz manifold.

\subsubsection{Fluids of matter.}
In General Relativity, a flow of particles is described by a fluid. :
\begin{greybox}
A \emph{dust} fluid of charged particles in a spacetime $(\M,\g)$ is a triple $\F=(\mu, e,X)$ where $\mu : \M\rightarrow \R^+$ and $e:  \M \rightarrow \R$ are smooth functions, and where $X$ is a future-directed timelike vector field on $\M$. We suppose furthermore that $\Div (\mu X)=0$ and  $\Div (e X)=0$. These last equations are called respectively energy and charge conservation laws. $\mu$ is called the baryonic number density, and $e$ the charge density.
\end{greybox}

\subsubsection{Electromagnetism.}
Let  $\F=(\mu, e,X)$ be a dust fluid of charged particles. We call $\vec \F :=\mu X$ the matter-energy flow, and $\vec J :=e X$ the electric current.
\begin{greybox}
An \defin{electromagnetic field} on  a spacetime $(\M,\g)$ is a differential 2-form $\w F \in \Lambda^2(\T\M)$.

If  $\F=(\mu, e,X)$ is a dust fluid of charged particles and $\w F$ is an electromagnetic field on a spacetime $(\M,\g)$, we say that $(\M,\g,\F,\w F)$ satisfies \defin{Maxwell equations} iff :
\begin{enumerate}
\item $\dd \w F=0$, i.e $\w F$ is closed,
\item $(\Div \w F)^\sharp = \vec J$. 
\end{enumerate}
\end{greybox}
\emph{Remarks :} 

1/: $\Div \vec J= \Div (\Div \w F)^\sharp =0$, so the charge conservation law is a consequence of 2/.

2/: As $\w F$ is closed, Poincaré lemma implies that it is locally exact. Therefore, on any simply connected open set $\U \subset \M$, there exists a 1-form $\w A$ on $\U$, called the \defin{electromagnetic potential}, such that $\dd \w A=\w F$. (Any other potential is then of the form $\w A +\dd f$ for a smooth function $f$ on $\U$).
\\

Let $\gamma :(-\epsilon, +\epsilon)\rightarrow \M $ be a curve such that $\dot \gamma (t)=-\mu^{\frac{1}{2}}X$ ; it represents the world-line of a particle of the flow. 
\begin{greybox}
$\gamma$ and $\w F$ satisfy the \defin{Lorentz law} if 
$$\forall t \quad \wnab_{\dot\gamma(t)} \dot\gamma(t)=e. ^e \w F(\dot\gamma(t)).$$
\end{greybox}
The particle is deviated from the geodesics $\wnab_{\dot\gamma(t)} \dot\gamma(t) =0$ by $\w F$.

\subsubsection{Energy-momentum tensors.}
On a spacetime $(\M,\g)$, let  $\F=(\mu, e,X)$ be a fluid of charged particles. Let also $\w F$ be an electromagnetic field.
Remember that the energy-momentum tensor of the matter-energy flow $\vec \F :=\mu X$ of the fluid is defined to be :
\begin{greybox}
$$\w T^m:= \mu X \ts X$$
\end{greybox}
where we identify this (2,0)-tensor with the corresponding (0,2)-tensor via musical isomorphism. In coordinates $(T^m)_{ij}=\mu X_i \, X_j$ if $X=X^i \partial_i$ for a coordinate frame $(\partial_i)$.
\begin{greybox}
Let $(\partial_i)$ be a coordinate frame where $\w F = (F_{ij})$ and $\g=(g_{ij})$. The energy-momentum tensor of the electromagnetic field $\w F$ is the (0,2)-tensor $\w T^F$ given by :
$$T^F_{ij} = F_{ik} F^k_{\ j} + \frac{1}{4} F_{kl} F^{kl} g_{ij}$$
$\w T^F$ is well-defined (i.e. independent of the choice of $(\partial_i)$), and symmetric. Furthermore, $trace (\, ^e \w T^F)=0$, and for any vector $X$ timelike or lightlike, $\w T^F (X,X) \geq 0$.
\end{greybox}
A justification for this tensor can be found in Eric Gourgoulhon's book : Special Relativity.

The link with the current $\vec J =e. X$ is then given by :
\begin{greybox}
If $(\M,\g, \F, \w F)$ satisfies the Maxwell equations, we have :
$$(\Div \w T^F)^\sharp = -\, ^e \w F (\vec J).$$
\end{greybox}

\subsubsection{Neutral fluids and Einstein equation.\\
Fundamental laws on fluids can be deduced from Bianchi's identity}
We show in this section  that conservation and motion equations for an electrically neutral fluid are simple consequences of the Bianchi identity $\Div \w G =0$.

In a spacetime $(\M, \g)$, families of particles are modeled by the flow lines of a vector field $X$, where $X$ is an everywhere unit timelike vector : $\g (X,X)\equiv -1$. Therefore, at each point $p$, the vector $X_p$ is the 4-velocity of the particle represented by the flow line of $X$ passing by $p$.

When the particles are not interacting, the energy-momentum tensor associated to such a family is given by $\w T := \mu   X\ts X$ where $\mu$ is a function $\mu  : \M \rightarrow \R$, $\mu\geq0$, called the \defin{energy density} of the fluid. Particles are present where $\mu  >0$. Such a model of particles is called a \defin{dust} matter fluid.

When one wants to consider non-gravitational interaction between the particles, one introduces a \defin{pressure} term $p$, also a function $p : \M \rightarrow \R$, satisfying some equation of state, and the energy-momentum tensor is then $\w T := \mu  X\ts X + p ( \g+ X\ts X).$
 Such families of particles are called \defin{fluids of matter}. 

Giving some conditions on the pressure $p$ leads to the fundamental definitions of \defin{perfect fluids} which are used to model stars or planets. Indeed, remember that a particle is modeled by a world-line in spacetime. Therefore, the flow of $X$ is to be considered as modeling the world-"tube" of the bundle of particles where $\mu  >0$, and for example the world-"tube" in spacetime of a star or planet.

The remarkable fact, giving very strong consistency to General Relativity, is then that applying the purely geometrical Bianchi identity, stating that the divergence of the  Einstein tensor vanishes, gives the equations of motion as well as the conservation laws for these families of particles.
\begin{greybox}
\begin{theorem}
If the spacetime $(\M,\g)$ satisfies the Einstein equation $\w G=\w T$ where the energy-momentum tensor is that of a perfect "dust" fluid $\w T=\mu  X\ts X$, then the Bianchi identity $\Div \w G=0$ gives the following two fundamental equations for matter:
\begin{itemize}
\item 1/ : Conservation law : $\Div(\mu  X)\equiv 0$. Using Stokes theorem, this is used to prove conservation of matter-energy ;
\item 2/ : Geodesic motion : when $\mu>0$, $\wnab_X X =0$. That is, \emph{the flow lines of massive particles are geodesics.} We recover the equivalence principle.
\end{itemize}
\end{theorem}
\end{greybox}
\begin{proof}
One just have to compute $\Div (\mu  X\ts X)$, which is also zero by Einstein equation and Bianchi identity. We have :
$$0=\Div (\mu  X\ts X)= (\Div (\mu  X)).X+\mu. \wnab_X X$$
But as $\la X,X \ra \equiv -1$, we have  $0=\wnab_X \la X,X \ra=2\la \wnab_X X, X\ra $. (Remember that we note $\g(X,Y):=\la X,Y \ra$). So taking the scalar product $\la X, \Div (\mu  X\ts X) \ra$, we get 
$$0= (\Div (\mu X)).\la X,X \ra +\mu  \la \wnab_X X , X\ra =-(\Div (\mu  X)).$$
But then, $\Div (\mu  X\ts X)$ reduces to $\mu .\wnab_X X$, which is therefore zero.
\end{proof}

In case of pressure, an analog proof using Bianchi identity also gives the state equation and the equation of motion. Remark : In the case of a perfect fluid, the fluid's curves are not
necessarily images of geodesics.

\subsubsection{Electromagnetism and Einstein equation.}
 For our relativistic model  $(\M,\g, \F, \w F)$, we now want to write Einstein equation : $\w G = \w T^m +\w T^F$.
 
 But Bianchi identity, $\Div \w G =0$ requires that $\Div (\w T^m +\w T^F)=0$. This will be a consequence of Maxwell equations and Lorentz law.
 
 Indeed, for our fluid $\F=(\mu, e,X)$ and its matter flow $\vec \F=\mu .X$ and its electric current $\vec J=e.X$, each flow line of $\vec \F$ represents a particle. For each of these particles, the Lorentz law is written : 
 $$\mu \wnab_X X =e (\, ^e \w F (X))= \, ^e \w F (\vec J).$$
 Now, we saw in the previous section that the Maxwell equations imply : \\
 $(\Div \w T^F)^\sharp = -\, ^e \w F (\vec J)$.\\
 Besides : $(\Div \w T^m)^\sharp=(\Div (\vec \F \ts X))^\sharp =(\Div \vec \F).X +\wnab_{\vec \F} X = (\Div \vec \F).X +\mu \wnab_X X$.\\
 But by our definition of a dust fluid, $\Div \vec \F =0$. \\
 So the Lorentz law implies : $\Div (\w T^m +\w T^F)=0$.

\subsubsection{Classical Axiomatic for Gravitation and Electromagnetism.}
We can now give the classical axiomatic for a relativistic model  $(\M,\g, \F, \w F)$ :
\begin{greybox}
Let $(\M,\g)$ be a spacetime, $\F=(\mu, e,X)$ be a dust fluid of charged particles, and $\w F$ an electromagnetic field on $\M$. As above, $\vec \F=\mu .X$ is the energy-matter flow, and $\vec J =e. X$ is the electric current. The Classical axiomatic for General Relativity and Electromagnetism, due to Faraday, Maxwell, Lorentz and Einstein, is :
\begin{enumerate}
\item \emph{Conservation laws :} $\Div (\mu X)=0$ and  $\Div (e X)=0$.
\item \emph{Maxwell equations :} $\dd \w F=0$ and $(\Div \w F)^\sharp = \vec J$.
\item \emph{Lorentz Law :} $\mu \wnab_X X =e (\, ^e \w F (X))$.
\item \emph{Einstein equation :}  $\w G = \w T^m +\w T^F$.
\end{enumerate}
\end{greybox}

This box does not present the minimum number of axioms, as these four equations are not independent.

In Particular : ( 2. + 4. + Bianchi ) $\Rightarrow$ ( 1. + 3.)

\section{Classical General Relativity as pure Geometry.}
We apply here the A-B-C frame of the introduction to classical 4-dimensional general relativity :

A/ Space-time is a Lorentzian manifold of 4 dimensions. (Observers,
proper time, and space seen by an observer are defined as usual). 

B/ We canonically define data based on the Lorentzian manifold's curvature tensor which will physically represent : density of energy, density of mass of a fluid, pressure of a fluid, unit vector of fluid curves, etc\ldots No physical object is added : there is only geometry. 

C/ No law is added. Bianchi's second identity gives mass conservation law (when appliable), the fact that for a perfect dust fluid, curves are geodesic, the equation verified by a perfect fluid, etc\ldots

Hence this identity gives us : an approximation of classical mechanic
(gravitation), big bang and big crunch for an isotrop and homogenous domain,
the study of spherical symmetry in space (Schwarschild) and therefore movements
of planets, light deviation, black holes.
We precisely find all general relativity applied to perfect fluids.

Unfortunately, this vision can not deal with electromagnetism. Indeed, even
though we can define in a canonical manner the energy-impulsion tensor
representing electromagnetism, we can not find a canonical definition for the
2-form of electromagnetism and Maxwell-Lorentz equations.
It is this precise problem that precursors Einstein, Weyl, Kaluza, Klein,
Rainich have attempted to solve. We'll get back to this point later on.
      
To summarize, the study of electrically neutral perfect fluid in general relativity can be
reduced to the study of Lorentzian manifolds. In other words, physical laws
regarding matter fluids are just translations of Riemannian geometry theorems.
However, it is not the case for electromagnetism which needs  the introduction
of an exact 2-form verifying  ``laws'', namely
Maxwell equations, in the space-time manifold, in order to have a
formal definition in general relativity.

The idea is that, in Einstein equation $\w G=\w T$, the curvature $\w G$ is the clear, well defined, mathematical object, whereas $\w T$ is the unclear, model-dependant, physical object. So we can decide that physical objects such as energy, mass, velocity of particles, are just peculiar aspects of the geometry of spacetime. Physical equations, motion or conservation for example, will then be geometrical facts such as Bianchi identity.
\\

\textbf{Example 1 : Scharzchild Geometry. }
Schwarzschild geometry does not requires the full Einstein equation, only the fact that empty spacetime is Ricci flat. It is then only based on the first geometrical postulate : Spacetime is a Lorentz manifold, admitting a 3-dimensional spherical symmetry (to be precisely defined) and Ricci flat.. It nevertheless gives fundamental modelization of gravitation around most astronomical objects. It is thus robust.
\\

\textbf{Example 2 : Geodesic motion for matter fluids as a geometric fact.}
As we just saw, a natural modelization of a bundle of free particles is a timelike flow, a fluid, giving rise to a natural 2-tensor $\rho X \ts X$ where $X$ is a unit timelike vector field and $\rho$ a positive function on $\M$. Postulating Einstein equation, we then decide that the Einstein curvature $\w G$ must be equal to  $\rho X \ts X$, and then we deduce geometrical facts, as we proved in the above theorem.

But we could see things in a purely geometrical way. Consider the endomorphism $^e \w G$ associated to Einstein curvature $\w G$ by musical isomorphisms. We could  \emph{define geometrically} a fluid as a domain of spacetime $(\M,\g)$ such that, at each point $p$, $^e \w G$ admits a 1-dimensional timelike eigenspace of dimension 1, and a 3-dimensional eigenspace associated to the eigenvalue $0$. Taking smoothly a unit eigen-vector $X(p)$ of the timelike eigenspace then produces a unit timelike vector field $X$. Considering the negative eigenvalue to be $-\rho (p)$ at each point, we \emph{define} $\rho(p)$ to be the energy-density, and the associated unit eigen-vector $X(p)$ to be the 4-velocity of the particle at $p$. It can then easily be shown that $\w G$ can be written $\w G= \rho X \ts X$. Then, as for the above theorem, applying Bianchi identity $\Div \w G =0$, we find  $\wnab _X X  =0$, that is, $X$ is a geodesic vector field, and its flow lines are geodesics. We recover equivalence principle as a purely geometric fact. 
\\

\textbf{Example 3 : Perfect fluid and electromagnetic field}

We consider a domain is spacetime where $^e \w G$ has the following properties at each point $x$:

1/: $^e \w G_x$ has an eigenvalue $-\mu <0$ of eigenspace $\mathcal E_{-\mu}$,
 $dim(\mathcal E_{-\mu})=1$ and spacelike.

2/: $^e \w G_x$ has an eigenvalue $\lambda_1$ of eigenspace $\mathcal
E_{\lambda_1}$, $dim(\mathcal E_{\lambda_1})=1$, such that
$\mathcal E_{\lambda_1} \perp_g \mathcal E_{-\mu}.$

3/: $^e \w G_x$ has an eigenvalue $\lambda_2$ of eigenspace $\mathcal
E_{\lambda_2}$, $dim(\mathcal E_{\lambda_2})=2,$ such that
$\mathcal E_{\lambda_2} \perp_g (\mathcal E_{\lambda_1} \oplus \mathcal
E_{-\mu})$  and such that $-\mu < \lambda_1 <\lambda_2 <\mu.$

This is equivalent to the existence of a $\g$-orthonormal base in which the matrix of $^e \w G$ is :
$$
\left( G^i_{\  j} \right) = \begin{pmatrix}
-\mu & 0&0&0\\
0 &  \lambda_1&0&0\\
0&0&\lambda_2&0\\
0&0&0&\lambda_2
\end{pmatrix}
\qquad \mbox{ with  } \mu >0 \mbox{ and } -\mu < \lambda_1 < \lambda_2 < \mu.
$$
Such domains physically represent \emph{the association of a perfect fluid and an
electromagnetic field.}

At each point $x$ we can then define unambiguously:
 \emph{The unit tangent vector $X_x$ to the fluid's
 curve} by the only unit vector in the orientation of $\mathcal E_{-\mu}.$
 \emph{The fluid's energy density at $x$ } by the positive real number
$\mu$ (sum of the fluid's energy density and the electromagnetic's
energy's density defined below).
 \emph{The fluid's energy density at $x$} by the positive real
number $\mu - \frac{1}{2}(\lambda_2 - \lambda_1).$
 \emph{The electromagnetic energy density at $x$} by the positive
real number $ \frac{1}{2}(\lambda_2 - \lambda_1).$
\emph{The fluid's pressure at $x$} by the real number $\frac{1}{2}(\lambda_1+\lambda_2).$
 \emph{The electromagnetic pressure at $x$} by the real number
 $\frac{1}{6}(\lambda_2 - \lambda_1)$. With these datas, we can define
the following tensor: \begin{center}$\w T =\frac1{2} (\lambda_2 - \lambda_1) X\ts X + \frac{1}{6} (\lambda_2 -
\lambda_1) (\g + X\ts X) + \Pi$,\end{center} or, in a coordinate frame, 
$T_{ij} = \frac1{2} (\lambda_2 - \lambda_1) X_i X_j + \frac{1}{6} (\lambda_2 -
\lambda_1) (g_{ij} + X_i X_j) + \Pi_{ij}$, where  $\Pi$ is a 2-tensor with trace 
equal to zero and such that $X_x \in Ker \, ^e \Pi_x$. $\Pi$ is called  \emph{the electromagnetic tensor in $x$}.

This tensor correspond to the classic 
energy-impulsion tensor of electromagnetism, but it does not  allow to
retrieve the electromagnetism 2-form $\w F$ canonically (and a fortiori
Maxwell's equations). Indeed, for a given symmetrical tensor $T_{ij}$, there
exists in general an infinite number of anti-symmetrical tensors $F_{ij}$ such
that $ T_{ij} = F_{ik} F^k_{\ j} + \frac{1}{4} F_{kl} F^{kl} g_{ij}.$
Therefore, it is not possible to retrieve classical electromagnetism,
({\it i.e} the $2$-form $\w F$ and Maxwell equations) with only 
the Lorentzian manifold's geometry as given here. However, we can wonder if the
energy-impulsion tensor $T_{ij}$ of electromagnetism is sufficient to describe
 physical reality, in particular a fluid's behavior
 (since, in the end, only fluids are physically observable).
 The answer is still no. The opposite would mean we could describe electromagnetic phenomenons without having to use the  $2$-form $\w F$, in other worlds
without using the electromagnetic field. It can be shown
that the knowledge of the tensor $T_{ij} $ alone can not lead to a physical
theory sufficiently deterministic, contrary to the classical theory of
electromagnetism in general relativity (which consists in introducing the
2-form $\w F$ with its energy-impulsion tensor in the
space-time, and postulating Maxwell equations).

\noindent \textbf{Therefore, in 4
dimensions, one can not describe electromagnetism using the
Lorentzian manifold's geometry alone.}

\section{Geodesic free fall in electromagnetic and gravitational fields. } 
As we saw above, the classical objects of electromagnetism cannot be obtained from the geometry of a 4-dimensional Lorentzian manifold. What can we try ? Obviously, if we want to obtain geometrically more objects, we need to enrich the geometry. Historically, one of the most famous method is due to Nordstrom, Kaluza and Klein : it consists in augmenting the dimension of the space-time manifold.

We shall see that starting with this idea and building upon the ideas presented in the introduction, we will be able to propose a unified geometrical setting for both gravitation and electromagnetism. This will be obtained by suppressing a requirement usually made in papers on the subject, that is imposing a Ricci flat metric on the 5-dimensional space-time manifold, requirement which is not justified from our view point. This will be explained in subsection 4.3.

The purpose of this section is to introduce our ideas in the simple case of "dust", that is, a flow of massive and charged particles, whose only interaction are due to gravitation and electromagnetism. We will give the most general case, general fluid in more-than-five-dimensional spacetime in the next section.

\subsection{Five dimensional space-time, "Small" dimension.}

We start by trying to set a natural modelization of "a small fifth dimension". The idea is to add, at every point of the classical 4-dimensional spacetime, an extra degree of freedom, modeled by a circle, and asking that this circle be "small". Mathematically, this is simply a \emph{fibration}. Indeed, the method originally proposed by Kaluza and Klein was to use a 5-dimensional fibre bundle structure over a 4-dimensional base representing classical space-time. 

\emph{The model for space-time is a 5-dimensional Lorentzian manifold $(\M,\g)$ equipped with a principal $S^1$-fibre bundle structure, ( $S^1$ being the circle), $\pi :\M\rightarrow \B\M$.}

Principal fibre bundle theory can be found in several textbooks. However, in the case of a $S^1$-bundle, the theory is much simpler; we therefore here present an elementary vision. 
\begin{greybox}
\begin{definition} \textbf{Kaluza-Klein spacetime :} 
\begin{itemize}
\item $(\M,\g)$ is a Lorentzian manifold of dimension 5, time-oriented, such that the Lie group $S^1$ acts freely and properly on $\M$. Therefore $\pi :\M\rightarrow \B\M$ is a $S^1$-fibration and $\B{\M}:=\M /S^1$ is a manifold of dimension 4. 
\item Furthermore, we suppose that, for the action of $S^1$, there exists a metric $\B \g$ on $\B\M$ turning  $\pi : (\M,\g) \rightarrow (\B{\M},\B{\g})$ into a Riemannian submersion such that $\B{\g}$ has the signature $(-,+,+,+)$ and such that $\forall x \in \B{\M}$, $\pi^{-1}(x)$ is spacelike. For $x\in\M$, we will note $S^1_x :=\pi^{-1}(\pi(x))$ the fiber at $x$.
\end{itemize}
\end{definition}
\end{greybox}
If we suppose that $vol_{\g}( \pi^{-1} (x))=cst$ on $\M$, it can be shown that the fibers $\pi^{-1}(x)$ are geodesics of $\M$. This will be obtained below, differently. 
To say that the "fifth" dimension is small, one just needs to set $vol_{\g } (\pi^{-1}(x))=\epsilon$, with $\epsilon$ small according to some physical reference. $\B\M$ can be thought as "classical" spacetime.

The main advantage of this geometry of spacetime is that it gives a natural normalized vector field, unique up to orientation, that will represent the electromagnetic potential :

\begin{greybox}
\begin{definition} \textbf{Electromagnetic potential :}

\begin{itemize}
\item By choosing an orientation on $S^1$, we define a vector field  $\, \w Y$ on $\M$, by setting that in each point $x$ of $\M$, $\w Y_x$ is the vector tangent to the fiber $\pi^{-1}(\pi(x))$ at $x$, such that $\g(\w Y,\w Y)=1$ and in the chosen orientation. $\w Y$ is called the \defin{electromagnetic potential} of the spacetime $\M$. 
\item We then consider the 1-form $\w Y^\flat$ associated to $\w Y$ by $\g$. We note $\w F=\dd (\w Y^\flat)$ the differential of $\w Y^\flat$ ; $\w F$ is the \defin{electromagnetic field} on $(\M,\g)$.
\item We define \defin{the horizontal space} $H_x$ at $x\in \M$ as being the subspace of  $\T_x (\M)$ $\g$-orthogonal to $\w Y_x$. $H_x$ is a 4-dimensional Minkowski space. It represents, locally,  "classical" spacetime at $x$. Note that $H_x$ is naturally isometric to $\T_{\pi(x)}\B\M$. We will sometimes write simply $H$ when not specifying the point $x$.
\item We suppose that $\w Y$ is a Killing vector field ; this means that $S^1$ acts on $\M$ by isometries. The flow generated by $\w Y$ is then an isometry field, and $\w Y$ satisfies Killing equation : $\nabla_iY_j+\nabla_jY_i=0.$
Supposing $\w Y$ to be a Killing vector field and of constant norm $\g(\w Y,\w Y)=1$ is sufficient to prove that it is necessarily geodesic:
$\wnab_{\w Y} \w Y=0$, (see below).
\item At last, we can also write, slightly abusing notations, $\g=\B{\g} + \w Y^\flat\otimes \w Y^\flat$. 
\end{itemize}
\end{definition}
\end{greybox}

The next proposition, easy to prove, shows why it is natural to suppose that $\w Y$ is a Killing vector field, when supposing that the compact dimensions are "small":
\begin{greybox}
\begin{proposition}
\textbf{Averaging the metric on $S^1$ :} Let $\w Y$ be tangent to the fiber $S^1_x$ and such that $g (Y,Y)=-1$ as above, but without supposing that $\w Y$ is a Killing vector field. Let $\sigma$ be the 1-parameter group of diffeomorphisms associated to the flow of $\w Y$. Define the "averaged" metric $\B \g$ by :
$$\forall x\in\M , \quad \B\g_x:=\frac{1}{\ell_x}\int_{t_0}^{t_0+\ell_x}(\sigma^*(t)\g)_x.dt $$
where $\ell_x$ is the length of $S^1_x$ relative to $\g$. ($\B\g_x$ does not depend on the choice of $t_0$ as $\sigma_x(.)$ is periodic, of period $\ell_x$.)  Then, $\B\g (\w Y,\w Y)=-1$, and $\forall s\in \R$, $\sigma^*(s).\B\g=\B\g$. That is, $\w Y$ is a Killing vector field for $\B\g$.
\end{proposition}
\end{greybox}

We have defined all of our mathematical setting. We now are going to show that it gives, under a very natural definition of fluid, seen as a geometrical type-domain of space-time, using only geometrical theorems, and not postulating any law, the Einstein-Maxwell-Lorentz equations as well as all the classical conservation laws.

From now on, we suppose that we are in the setting given in the two definitions above.

\subsection{Matter fluids in 5-dimensional spacetime.}
We note $\w G=\w{Ric}-\frac{1}{2}\w S.\g$ the Einstein curvature. $^e\w G$ is the associated endomorphisms field. 
\\

In 4-dimensional spacetime, a fluid of matter (electrically neutral) is a domain where the Einstein-Ricci curvature can be written $\w G= \mu X\ts X+P$, where $X$ is a unit timelike vector field and $P$ a matrix such that $^eP(X)=0$. As we saw in the previous section, if we want to insist on the geometrical aspect of the definition, a fluid is a domain where $^e\w G$ possesses a timelike 1-dimensional eigenspace of which $X$ is a eigenvector. Dust is a fluid where $P=0$, i.e. $\w G= \mu X\ts X$.
\\

Our focus will now be on a natural extension of the definition of a fluid as a domain of  Kaluza-Klein spacetime whose Einstein curvature possesses a 1-dimensional timelike eigenspace. The definition of a fluid will just be slightly modified to require that the 1-dimensional timelike eigenspace of $^e \w G$ should have a timelike $\g$-orthonal projection on the 4-dimensional subspace modelizing classical space-time, (the fluid is not "flowing along the fifth dimension").

Suppose we are given, on an open subset of $\M$ where $\w Y$ is defined, a vector field $X_0$, timelike, of norm $\g(X_0,X_0)=-1$, and orthogonal at each point of $\M$ to $\w Y$, $(X_0)_x \,\bot_{\g}\, \w Y_x$. This vector field represents a family of observers. We recall that $H_x=\w Y_x^{\bot}$ is the horizontal space. 

We now define $^e \w G_H$, the endomorphisms field on the horizontal subspaces $H_x$, defined by $^e \w G_H=pr_H \circ (^e \w G_{|H})$, where for $x \in \M$, $(pr_H)_{|x}$ is the orthogonal projection of $\T_x \M$ on $H_x$. This tensor will be very important to define fluids.\\

The basic idea to define a fluid of matter is the following : 

 \emph{A fluid is a domain of $\M$ where there exists a naturally defined timelike vector field. More precisely, at least to begin with, a fluid is a domain of $\M$ where $^e\w G_H$ admits in each point a eigenspace of dimension 1, timelike, and orthogonal to $\w Y$.} 
\begin{greybox}
\begin{definition} 
 A domain $\Omega \subset \M$ is a \defin{perfect charged matter fluid domain} if and only if at each point :  $^e\w G_H$ has a timelike 1-dimensional eigenspace $E_{-\mu}$ of eigenvalue $-\mu<0$, and \\
 $^e\w G(\w Y) \in<\w Y,E_{-\mu}>$, the vector space generated by $\w Y$ and $E_{-\mu}$.
 
 This is the case if and only if its Einstein curvature tensor can be written 
 $$\w G=\mu X \otimes X + \alpha \w Y \otimes \w Y+P$$
with the condition that, at each point $x$, $pr_H(X)$ is a basis for a timelike 1-dimensional eigenspace of $^e\w G_H$ of eigenvalue $-\mu<0$, and $P$ is a matrix such that 
$^eP(X)=\, ^eP(\w Y)=0$. 

If $P=0$ (which means a fluid with no pressure), then $X$ is unique for the decomposition\\
 $\w G=\mu X \otimes X + \alpha \w Y \otimes \w Y$ (once a time orientation is chosen).
 \end{definition}
\end{greybox}

\emph{In this section, we want to specialize on "electric dust", that is, a model of a flow of particles  whose only interactions are due to gravitation and electromagnetism :}
\begin{greybox}
\begin{definition}
 A domain $\Omega \subset \M$ is a \defin{dust charged matter fluid domain} if and only if its Einstein curvature tensor can be written 
 $$\w G=\mu X \otimes X + \alpha \w Y \otimes \w Y$$
with the condition that, at each point $x$, $pr_H(X)$ is a basis for a timelike 1-dimensional eigenspace of $^e\w G_H$ of eigenvalue $-\mu<0$. $X$ is then unique for this decomposition.

\textbf{Associated "classical" data : } For such a perfect fluid without pressure, there is a unique decomposition (once a time orientation is chosen) :
$$\w G= \mu X_0 \otimes X_0 -e(X_0 \ts \w Y+\w Y \ts X_0)+\gamma \w Y \ts  \w Y,$$
where $\g(X_0,X_0)=-1$ and $X_0 \bot \w Y$. $\mu$ is called mass density, $e$ the charge density. These are canonically given by : $$\mu=-\w G(X_0,X_0)$$ $$e=\w G(\w Y,X_0)$$ $$\gamma= \w G(\w Y,\w Y)$$
 We then have 
 $$X=X_0-\frac{e}{\mu}\w Y$$
The vector field $X=X_0- \frac{e}{\mu}\w Y$ is called the vector field of the fluid, and the associated flow, the flow of the fluid.
The vector field $X_0$ will be called the \emph{apparent, or visible,} field of the fluid, and the associated flow, the \emph{apparent, or visible,} flow. Note that at each point, by definition, $X_0(x)\in H_x$.
 \end{definition}
 \end{greybox}

\subsection{Do not kill Ricci.}
Our geometrical setting for 5-dimensional spacetime produces a natural vector field $\w Y$, and from this, a 2-form $\w F$ that we identify with the electromagnetic field 2-form. This is now the point where we depart from the articles we know about. In these, it is always considered that 5-dimensional space-time must be Ricci-flat. However, in the frame of Kaluza-Klein theory, this implies with the usual hypothesis made, that $|\w F|_g=0$, which contradicts the requirement of electromagnetism. Thus nothing can be obtained this way.

However, from our point of view, there is no reason to ask for the Ricci curvature to be zero. We can see this Ricci=0 requirement as a way to consider that matter is "added" to space-time; geometry cames next. From our point of view, there is only geometry, thus curvature ; matter is only an aspect of geometry.
 
Relieving this "Ricci=0" requirement, we will see that the known Kaluza-Klein formulae give the classical Einstein and Maxwell-Lorentz equations, that is gravity and electromagnetism, using only geometrical theorems or formulae. 

We will start by giving very general equations for space-time dynamics as seen by a family of timelike observers. We will then see that if these observers are linked to a massive charged fluid, defined in a purely geometrical way, and if they can only see what is happening on their 4-dimensional space-time, they will recover the classical equations of physics.

\subsection{Geodesic free fall in electromagnetic and gravitational fields.}

It is very important to note that we have never mentioned any kind of energy-momentum tensor. The point is that from our point of view, this concept has no meaning. Indeed, let us review our frame of ideas :

A/: Space-time is a five dimensional semi-Riemannian manifold satisfying definition 1

B/: Instead of defining an energy-momentum tensor, we caracterize a domain of space-time by a geometric type. We \emph{then} define physical concepts by geometric caracteristics of curvature.

C/:  Physical equations are projection of the Bianchi identity $\Div \w G=0$ on the 4-dimensional subspace $H=\w Y^{\bot}$ modelizing our classical 4-dimensional space-time.
\\

We also want to consider the idea of \emph{free fall in an electrogravitational field.}. Indeed, one of the cornerstone of general relativity is the equivalence principle. It is expressed mathematically by the hypothesis that free particles follow time-like geodesics of space-time. For perfect fluids without pressure (dust), whose energy-momentum tensor is $\mu X_0\otimes X_0$, it is expressed by the fact that the vector field $X_0$ associated to the flow lines is a geodesic vector field. We recalled at the beginning of this chapter that this fact is obtained by applying the Bianchi identity to this tensor when considered as the Einstein curvature; once again it is just a purely geometrical fact. This can be summarized by the following theorem, which is a rewriting of the axiomatic for gravitation and electromagnetism given in 2.0.6 :
\begin{theorem}
(Einstein, 1916) Space-time is a 4-dimensional Lorentzian manifold $(\M,\g)$. A perfect dust fluid is a domain $\Omega$ of space-time whose Einstein curvature is of the form $\w G=\mu X_0 \otimes X_0$, $X_0$ being a timelike vector field. The Bianchi identity implies that $X_0$ is a geodesic vector field, and that $\Div (\mu X_0)=0$. To modelize electromagnetism, one then add a closed 2-form $\w F$, a function $e: \Omega \rightarrow \mathbb{R}$, and postulate the Lorentz and second Maxwell equations, as well as the conservation of charge $\Div (eX_0)=0$. (In fact, given $\w F$ and $e$, it is sufficient to postulate the first and second Maxwell equations, Bianchi giving the Lorentz law.)
\end{theorem}

Considering the inclusion of electromagnetism in the geometrical frame of space-time, it would be satisfactory to extend the equivalence principle to our five-dimensional setting. In the case of our charged dust fluid above (definition 8), the equation of movement on classical four-dimensional space-time $\w Y^{\bot}=H$ is $$\mu \wnab_{X_0}X_0 =e.\,^e\w F(X_0).$$ 
This is Lorentz law and $X_0$ is of course not geodesic in general. However, we would like it to be the "trace" on classical space-time, that is the projection on $H$, of a geodesic trajectory in five dimensions. Obviously this has to involve movement along the "small" fifth dimension. 
Our main result is then the following, to be compared to the axiomatic for gravitation and electromagnetism given in 2.0.6 :

\begin{greybox}
\begin{theorem}\textbf{Dynamics of charged dust.}
For the domain of a perfect charged fluid without pressure where $\w G$ and its associated classical data are written :
$$\w G=\mu X \otimes X + \alpha \w Y \otimes \w Y= \mu X_0 \otimes X_0 -e(X_0 \ts \w Y+\w Y \ts X_0)+\gamma \w Y \ts  \w Y,$$
the Bianchi identity gives:
\begin{itemize}
\item Conservation Laws: $X_0(\frac{e}{\mu})=\Div (\mu X_0)=\Div (eX_0)=0$ 
\item Maxwell equations :  $\dd \w F=0$ and $(\Div \w F)^\sharp =2eX_0-(2\gamma +\w S_g)\w Y$
\item Free Fall : $X=X_0-\frac{e}{\mu}\w Y$ is a geodesic vector field.
\item Lorentz equation: $\mu \wnab_{X_0} X_0=e. ^e\w F(X_0)$. This is just free fall read on $H$.
\end{itemize}
When projected on the "classical" 4-dimensional space-time $H=\w Y^\bot$, these equations are the classical equations of physics. 

Note that Lorentz equation is obtained from the geodesic motion of $X=X_0-\frac{e}{\mu}\w Y$ by developing $\wnab_X X=0$ and writing (projecting) this equation on the horizontal space $H=\w Y^{\bot}$, noting that $\wnab_{X_0} X_0 \bot \w Y$, which means that $\wnab_{X_0} X_0\in H$. We therefore see that : 
\begin{center}
 \emph{Free fall for $X$ is equivalent to Lorentz equation for $X_0$}. 
 \end{center}
\end{theorem}
\end{greybox}
  (Remember that the first Maxwell equation, $\dd \w F=0$, is always obvious as we set $\w F=\dd \w Y^\flat$.)
 \begin{greybox}
\textbf{Dynamics of a perfect charged fluid with pressure :} \\
A perfect charged fluid with pressure is a domain whose curvature is of the form $$\w G=\mu X \otimes X + \alpha \w Y \otimes \w Y+P$$
for some matrix $P$ such that $^eP(X_0)=\, ^eP(\w Y)=0$. $P$ is called the pressure/constraint tensor. The choice of $P$ corresponds to the choice of a state equation for the fluid. Then, the Bianchi identity gives :
$$(\Div \w F)^\sharp = 2e. X_0 - (2 \gamma+\w S_{\g}).Y $$
$$\Div (\mu X_0) -\la X_0, (\Div P)^\sharp \ra=0$$
$$\Div (eX_0)=0$$
$$ \mu \wnab_{X_0}X_0 -e.\,^e\w F(X_0)+ (\Div P)^\sharp =0 $$
\end{greybox}

 \begin{proof}
The proof will be given below, but here is the scheme for a charged dust fluid :

First some properties of $\w Y$ are established. Then, considering $\Div  \w G$ as a vector, i.e. identifying $\Div \w G$ and $(\Div \w G)^\sharp$, and noticing that $\Div \w G =0$ by Bianchi identity, one compute:

\begin{itemize}
\item $\g(\Div \w G, \w Y)$, this will be charge conservation law.
\item $\g(\Div \w G, X_0)$, this will be mass (baryonic number) conservation law.
\item $pr_H (\Div \w G)$, this be the equation of motion. In the case of dust $pr_H (\Div \w G)=\Div \w G$, so the equation of motion is simply $\Div \w G=0$.
\end{itemize}
Then, $\Div \w F=\Div (\dd\w Y^\flat)$ is computed, which gives the second Maxwell law, the first, $\dd\w F=0$, being obvious as $\w F=\dd (\w Y^\flat)$.

Finally, to prove that the flow of $X$ is geodesic, we simply compute $\wnab_X X$, noticing that $X_0 (\frac{e}{\mu})=0$ ; this leads to $\wnab_X X=0$.

Considering a fluid with pressure adds some technicalities, but the proof remains essentially the same.
 \end{proof}

To end this section, one can mimic some classical definitions : 
\begin{enumerate}
\item An observer of $(\M,\g)$ is a timelike curve $\gamma : I \rightarrow M$.
\item The space-time seen by $\gamma$ at $x=\gamma (t)$ is $\w Y_x^{\bot}$.
\item The full space seen by $\gamma$ at $x=\gamma (t)$ is $\dot \gamma  (t) ^{\bot}$.
\item The classical space seen by $\gamma$ at $x=\gamma (t)$ is $<\w Y_{\gamma (t)},\dot \gamma (t)>^{\bot}$, (i.e $H_{\gamma (t)}$). 
\item A  \emph{classical}, or \emph{galilean}, observer, is a timelike curve $\gamma$  which is horizontal, i.e. $\dot \gamma (t) \bot \w Y_{\gamma (t)}$ for all $t$. If such an observer can only see 4 dimensions and not the fifth carried by $\w Y$, then his "measure process" are projections on his horizontal space-time, $H_{\gamma (t)}$.
\end{enumerate}

\emph{Remark : The factor 2 in front of $e.X_0$ in the second Maxwell equation, for dust and for a fluid with pressure, is just a matter of convention : Replace the einstein tensor $\w G$ in the given definitions of fluid by $\Tld {\w G}= 2\eta^{-1}.\w G$ to get $\Div \w F= \eta .e. X_0 + (\eta . \gamma+\w S_g).\w Y $, the other equations being unchanged. See the computation of $\Div \w F$ below.}

\subsection{Some technical preliminaries.}
We consider Kaluza-Klein spacetime $\M$, a $S^1$-fiber bundle $\pi:\M\rightarrow \B\M$ over a 4-dimensional Lorentz manifold $\B \M$. $\w Y$ is the naturally defined unit spacelike vector tangent to the fibers $\pi^{-1}(\B x)$, and $\w F:=\dd (\w Y^\flat)$. The vector field $\w Y$ is supposed  to be a Killing vector field.

We recall some notations. We note $\w G=\w{Ric}-\frac{1}{2}\w S.\g$ the Einstein curvature. $^e\w G$ is the associated endomorphisms field. We note $^e \w G_H$ the endomorphisms field on the horizontal spaces $H_x$ defined by $^e\w G_H=pr_H \circ (^e\w G_{|H})$, where for $x \in \M$, $(pr_H)_{|x}$ is the orthogonal projection of $\T_x \M$ on $H_x$.
\\

Let us consider the domain of a dust charged fluid where $\w G=\mu X \otimes X + \alpha \w Y \otimes \w Y$  and let $X_0=pr_H(X)$ be a basis for a timelike 1-dimensional eigenspace of $^e\w G_H$ of eigenvalue $-\mu<0$, with $\g(X_0,X_0)=-1$. 

We start by important consequences of the fact that $\w Y$ is Killing, and important facts on the horizontal space $H_x$  :
\begin{greybox}
\begin{proposition}
Regarding $\w Y$ and the horizontal spaces $H_x$ :
\begin{enumerate}
  \item $\wnab_{\w Y} \w Y=0$, that is, $\w Y$ is geodesic.
   \item $\Div \w Y =0$.
  \item A 1-dimensional, timelike, eigenspace of $^e\w G_{H_x}$ is necessarily unique in $H_x$.
  \item $X_0$ is invariant under the flow of $\w Y$, i.e. the Lie derivative of $X_0$ along $\w Y$ vanishes: $\Liesymbol_{\w Y} X_0=[X_0,\w Y]=0$. Therefore  $\wnab_{\w Y} X_0=\wnab_{X_0} \w Y$.
  \item $\w Y (e)=\w Y(\mu )=\w Y (\gamma)=0$.
  \item The horizontal space $H_x$ remains horizontal under the flow of $\w Y$.
  \item The 2-form $\w F:=\dd (\w Y^\flat)$  satisfies $\w F (V,W)= 2\la \wnab_V \w Y,W\ra$. So, the associated endomorphism $^e\w F$ is $V\mapsto 2\wnab_V \w Y$.
 \end{enumerate}
\end{proposition}
\end{greybox}
\begin{proof}
1 : Because $\w Y$ is Killing, we have for any vector field $V$, 
$$\la \wnab_V \w Y,\w Y\ra= - \la V, \wnab_{\w Y}\w Y\ra.$$
Now, $\la \w Y,\w Y\ra \equiv 1$, so for any $V$, $\wnab_V \la \w Y,\w Y\ra=0$. Therefore 
\begin{align}
0 &=\wnab_V \la \w Y,\w Y\ra \nonumber \\
&=2\la \wnab_V \w Y,\w Y\ra \nonumber \\
&=-2\la V, \wnab_{\w Y}\w Y\ra \nonumber
\end{align}
That is, for any $V$, $\la\wnab_{\w Y}\w Y , V \ra=0$. Thus $\wnab_{\w Y}\w Y=0$.
\\

2 : Pick a frame $(E_i)$ around some point $x\in\M$. By definition 
$$\Div \w Y := \sum \epsilon_i \la \wnab_{E_i} \w Y, E_i\ra.$$
But as $\w Y$ is Killing, $\la \wnab_{E_i} \w Y, E_i\ra=-\la E_i, \wnab_{E_i} \w Y\ra$. So the above sum is zero.
\\

3: Let $X_0$ and $X_0 '$ be eigenvectors of  $^e\w G_{H_x}$ associated to eigenvalues $\lambda$ and $\lambda '$ respectively. Because $\w G$ is symmetric and $X_0 , X_0 '$ are in $H$, we have 
$$\la ^e\w G_{H_x}(X_0), X_0 ' \ra =\w G (X_0,X_0 ')=\la X_0, ^e\w G_{H_x}(X_0 ') \ra.$$
But then :
\begin{align}
\la ^e\w G_{H_x}(X_0), X_0 ' \ra  & =\w G (X_0,X_0 ') \nonumber \\
&=\lambda \la X_0,X_0 '\ra \nonumber \\
&=\la X_0, ^e\w G_{H_x}(X_0 ') \ra \nonumber \\
&=\lambda ' \la X_0,X_0 '\ra  \nonumber 
\end{align}
As $H_x$ is a Lorentz vector space, and as $X_0$ and $X_0'$ are both timelike and non-null, $ \la X_0,X_0 '\ra\not =0$. So $\lambda=\lambda'$, and $X_0=c.X_0'$ for some non-null constant $c$.
\\

4 : Let $\psi_t$ be the flow of $\w Y$. Because $\w Y$ is Killing, $\psi_t$ is an isometry for any $t$. For $x\in\M$, let us note $S^1_x :=\pi^{-1}(\pi(x))$ the fiber at $x$. By definition $\w Y$ is invariant by $\psi_t$, so $\T_xS^1_x$ is stable, i.e. $\T_x \psi_t (\T_x S^1_x)=\T_{\psi_t(x)}(S^1_{\psi_t(x)})$. (Here, $\T_x \psi_t$ is the differential of the map $\psi_t$ at $x$, and $\T_{\psi_t(x)}(\N)$ is the tangent space of the submanifold $\N$ at the point $\psi_t (x)$.)

$\T_x \psi_t$ is an isometry, and $H_x=\w Y_x^\bot$. Therefore we also have 
$\T_x \psi_t(H_x)=\T_{\psi_t(x)}(H_{\psi_t(x)})$.

Then, because the 1-dimensional, timelike, eigenspace of $^e\w G_{H_x}$ is unique in $H_x$, and because $X_0$ is normalized, we have $\T_x \psi_t (X_0(x))=\pm X_0(\psi_t(x))$. Now, fix a point $p$ and let $V$ be a future-pointing timelike vector field. Suppose, for example, that $X_0$ is future pointing in a neighborhood of $p$ : $\la X_0(x),V(x)\ra <0$ for all $x$ near $p$.  As $\T_x \psi_t (X_0(x))$ remains timelike, for every $t$ and every $x$, we  have $\la \T_p \psi_t (X_0(p)), V(\psi_t(p))\ra \not =0$ for every $t$ near $0$. But $t\mapsto \la \T_p \psi_t (X_0(p)), V(\psi_t(p))\ra$ is continuous and $<0$ for $t=0$ ; it is therefore $<0$ for every $t$. Thus  $\T_p \psi_t (X_0(x))= X_0(\psi_t(p))$ for every $t$. This shows that $X_0$ is invariant by $\psi_t$, and so $\Liesymbol_{\w Y} X_0=[X_0,\w Y]=0$.
\\

5 : $e$, $\mu$ and $\gamma$ are defined by $\mu=-\w G(X_0,X_0)$, $e=\w G(\w Y,X_0)$ and $\gamma= \w G(\w Y,\w Y)$. We just proved that $X_0$ is invariant by the flow of $\w Y$, and so is  of course $\w Y$. The flow of  $\w Y$ being generated by isometries, $\w G$ is also invariant.  Thus the result.
\\

6 : This was proved with point 4.
\\

7: We want to compute $\w F:=\dd (\w Y^\flat)$. For this we will use the fact that, by definition, $\w Y^\flat (V)=\la \w Y, V\ra$, and the following formulae from chapter 3 :
$$\dd (\w Y^\flat)(V,W)=(\wnab_V \w Y^\flat)(W)-(\wnab_W \w Y^\flat)(V).$$
Now, 
\begin{align}
(\wnab_V \w Y^\flat)(W)&=V(\w Y^\flat(W))-\w Y^\flat (\wnab_V W) \nonumber \\
   &=V(\la W,\w Y \ra)-\la \wnab_V W,\w Y \ra \nonumber
\end{align}
Symmetrically, $(\wnab_W \w Y^\flat)(V)=W(\la V,\w Y \ra)-\la \wnab_W V,\w Y \ra$. So
$$\dd (\w Y^\flat)(V,W)=V(\la W,\w Y \ra)-W(\la V,\w Y \ra)-\la [V,W],\w Y\ra.$$
Then, writing $V(\la W,\w Y\ra)=\la \wnab_V W, \w Y\ra+\la W, \wnab_V \w Y\ra$, and using the symmetry of the connection : $[V,W]=\wnab_V W -\wnab_W V$, we get :
$$\dd (\w Y^\flat)(V,W)=\la \wnab_V \w Y , W \ra-\la \wnab_W \w Y , V \ra.$$
We now fundamentally use the fact that $\w Y$ is a Killing vector field, therefore having : 
$$\la \wnab_W \w Y , V \ra=-\la \wnab_V \w Y , W \ra.$$
 This gives 
$$\dd (\w Y^\flat)(V,W)=2\la \wnab_V \w Y , W \ra$$
which is equivalent to $^e(\dd \w Y^\flat)=2\wnab \w Y$.

\end{proof}

It follows easily some geometric properties of $\w Y$ and $X_0$. Let us call a vector field $V$ \defin{horizontal} if $V_x \in H_x$ for all $x$.
\begin{greybox}
\begin{proposition}
Regarding $X_0$ and $\w Y$, we have :
\begin{enumerate}
\item  $\la \wnab_{X_0} X_0, X_0 \ra =\la \wnab_{X_0} \w Y,\w Y\ra=\la \wnab_{\w Y} X_0, X_0\ra=0$.
\item $\la \wnab_{\w Y} X_0,\w Y\ra=0$.
\item $\la \wnab_{X_0} \w Y,X_0 \ra =0$ and $\la \wnab_{X_0} X_0,\w Y\ra=0$, so in particular $\wnab_{X_0} X_0$ is horizontal.
\end{enumerate}
\end{proposition}
\end{greybox}
Point 1/ is an easy consequence of the fact that $X_0$ and $\w Y$ are normalized. Point 2/ is a consequence of the fact that $\w Y$ is geodesic. Point 3/ is a consequence of the fact that 
$\Liesymbol_{\w Y} X_0=[X_0,\w Y]=\wnab_{\w Y} X_0-\wnab_{X_0} \w Y=0$. Furthermore, point 3 has the nice and important interpretation that the movement of the apparent fluid, the flow of $X_0$, is entirely determined on the "classical" horizontal 4-space $H$, as $X_0$ and $\wnab_{X_0} X_0$ belong to $H$.
\begin{proof}
1/ $\la X_0, X_0 \ra \equiv 0$, so $0=\wnab_{X_0} \la X_0, X_0 \ra =2 \la \wnab_{X_0} X_0, X_0 \ra$. The other equalities are obtained in the same manner.
\\

2/ $\la X_0, \w Y \ra \equiv 0$, so $0=\wnab_{\w Y} \la X_0, \w Y \ra =\la \wnab_{\w Y} X_0,\w Y\ra +\la X_0,\wnab_{\w Y}\w Y\ra$. But $,\wnab_{\w Y}\w Y=0$ as $\w Y$ is geodesic.
\\

3/ We saw in the previous proposition that $\Liesymbol_{\w Y} X_0=0$, so $\wnab_{\w Y} X_0 = \wnab_{X_0} \w Y$. Therefore $\la \wnab_{X_0} \w Y,X_0 \ra =\la \wnab_{\w Y} X_0,X_0 \ra =0 $ by point 1/. But now, $\la X_0,\w Y\ra\equiv 0$, so \\
$0=\wnab_{X_0}\la X_0,\w Y\ra= \la \wnab_{X_0} X_0,\w Y\ra+\la \wnab_{X_0} \w Y,X_0 \ra=\la \wnab_{X_0} X_0,\w Y\ra$.
\end{proof}

\subsection{Proof of the theorem on the dynamic of charged dust.}
To keep things simple in this section, we treat only the case of dust. The equations for a fluid with pressure are obtained with exactly the same proof, the presence of the pressure/constraint matrix $P$ introducing no difficulties. This case, in more than 5 dimensions, will be addressed in the next section.  
\subsubsection{Conservation laws and equation of motion.}
We have $\w G= \mu X_0 \otimes X_0 -e(X_0 \ts \w Y+\w Y \ts X_0)+\gamma \w Y \ts  \w Y$. By Bianchi identity, $\Div \w G =0$.
If we compute $\Div \w G$ using
$\Div (f.X\otimes Y):=c(\wnab(f.X\otimes Y))= (\Div (f.X)).Y+f.\wnab_X Y$
and $\Div (f.X\otimes Y)=(\wnab_X f). Y+ (f .\Div X).Y+f.\wnab_X Y$, identifying  by musical isomorphism $\Div \w G$ with the vector $(\Div \w G)^\sharp$, we have :
\begin{align}
\Div \w G =&[\Div (\mu X_0)]X_0 +\mu \wnab_{X_0}X_0 \nonumber \\
& -[\Div (e X_0)]\w Y -e  \wnab_{X_0}\w Y \nonumber \\
& -\w Y(e).X_0-e(\Div \w Y)X_0-e\wnab_{\w Y}X_0  \nonumber \\
& +\w Y(\gamma)+\gamma(\Div \w Y)\w Y+\gamma \wnab_{\w Y} \w Y \nonumber
\end{align}

But from the preliminary results, $\Div \w Y=0$, $\wnab_{\w Y} \w Y=0$, and $\w Y(e)=\w Y(\mu)=\w Y(\gamma)=0$. So
$$\Div \w G= [\Div (\mu X_0)]X_0 +\mu \wnab_{X_0}X_0  -[\Div (e X_0)]\w Y-e  \wnab_{X_0}\w Y-e\wnab_{\w Y}X_0.$$
Preliminary results also included that : 
$\la X_0,\wnab_{X_0}\w Y\ra=\la \w Y, \wnab_{X_0}\w Y\ra=\la X_0,\wnab_{X_0}X_0 \ra=0$.
Therefore, taking scalar product, and as $\Div \w G =0$, we obtain the conservation equations:
$$0=\la \Div \w G, X_0\ra = -\Div (\mu X_0)$$
$$0=\la \Div \w G, \w Y\ra =-\Div (e X_0)$$

For the equation of motion, Lorentz equation, note that we also saw that \\
$\wnab_{X_0}\w Y=\wnab_{\w Y}X_0$ as $[X_0,\w Y]=0$, and that $^e\w F= 2\wnab_V \w Y$. So finally :
$$\Div \w G= \mu \wnab_{X_0}X_0 -e.^e\w F (X_0)$$
which is zero by Bianchi identity, therefore giving the motion equation $ \mu \wnab_{X_0}X_0 =e.^e\w F (X_0)$. Remember that $\wnab_{X_0}X_0$ is horizontal, so this last equation is integrally written in $H$.

\subsubsection{Second Maxwell equation.}
It is obtained by computing $\Div \w F$. We shall consider the computation in some open set $\U$ of $\M$ where exists some frame field $(E_i)$. We will use the following formulas, see e.g. [O'Neill] :
$$\Div \w F (V)=\sum \epsilon_i (\wnab_{E_i}\w F)(E_i,V)$$
$$\Div \w Y =\sum \epsilon_i \la \wnab_{E_i}\w Y,E_i\ra $$
$$ \w {Ric} (V,\w Y) := \sum \epsilon_i . \la \w R (E_i, V) \w Y, E_i \ra ,$$
the sums being from $1$ to $dim\M=5$ and $V$ being some vector field in $\U$. We will also use the fact from preliminary results that $$\w F (V,W)= 2\la \wnab_V \w Y,W\ra$$
and the fact that for any vectors $V,W$, $\la \wnab_V \w Y, W\ra =-\la V,\wnab_W \w Y\ra$ as $\w Y$ is a Killing vector field, which is also the fact that the 2-form $\w F$ is antisymmetric.\\
First : $\wnab_{E_i} \w F (E_i,V)=E_i(\w F (E_i,V))-\w F(\wnab_{E_i} E_i,V)-\w F(E_i,\wnab_{E_i} V) $\\
Then : $\w F (E_i, V)=2\la \wnab_{E_i} \w Y, V\ra=-2\la \wnab_V \w Y,E_i\ra$ as $\w Y$ is Killing.\\
So, using again the antisymmetry of $\w F$ : \\
$E_i (\w F (E_i, V))=-2\la \wnab_{E_i} \wnab_V \w Y, E_i\ra-2\la \wnab_V \w Y, \wnab_{E_i} E_i\ra$\\
$-\w F (\wnab_{E_i} E_i, V)=+ 2\la \wnab_V \w Y, \wnab_{E_i} E_i\ra$\\
$-\w F(E_i, \wnab_{E_i} V)=-2\la \wnab_{E_i}\w Y, \wnab_{E_i} V\ra =+2\la \wnab_{\wnab_{E_i}V} \w Y, E_i\ra  $\\
Therefore : $$\Div \w F (V)= -2 \sum \epsilon_i \la \wnab_{E_i} \wnab_V \w Y -\wnab_{\wnab_{E_i}V} \w Y, E_i\ra$$
Now by definition of the curvature :
$$\wnab_{E_i}\wnab_V \w Y-\wnab_{\wnab_{E_i}V} \w Y=\wnab_V \wnab_{E_i} \w Y-\wnab_{\wnab_V E_i}\w Y 
+\w R (E_i,V)\w Y$$
So : $\Div \w F (V)= -2 \sum \epsilon_i \{ \la \w R (E_i,V)\w Y,E_i\ra-2\la\wnab_V \wnab_{E_i} \w Y, E_i\ra +2\la \wnab_{\wnab_V E_i}\w Y, E_i\ra \}  $
\\

But as $\w Y$ is Killing, $2\la \wnab_{\wnab_V E_i}\w Y , E_i\ra=-2 \la \wnab_{E_i} \w Y,\wnab_V E_i \ra $, \\ 
and furthermore 
$\la \wnab_V \wnab_{E_i} \w Y,E_i\ra+\la \wnab_{E_i} \w Y,\wnab_V E_i \ra =V(\la  \wnab_{E_i} \w Y, E_i\ra)$,  so we obtain :
\begin{align}
\Div \w F (V) &=-2 \sum \epsilon_i \{ \la \w R (E_i,V)\w Y,E_i\ra-2.V(\la  \wnab_{E_i} \w Y, E_i\ra)  \} \nonumber \\
 & = -2\, \w{Ric}(V,\w Y)-2V(\Div \w Y) \nonumber
\end{align}
and as $\Div \w Y =0$, we obtain the important formula :
$$\Div \w F (V)= -2\, \w{Ric}(V,\w Y).$$
Now $\w G=\w{Ric}-\frac{1}{2}\w S.\g$, so $\Div \w F (V)=-2\, \w G (V,\w Y)-\w S\la V, \w Y\ra$. \\
We now use our definition of a fluid : $\w G= \mu X_0 \otimes X_0 -e(X_0 \ts \w Y+\w Y \ts X_0)+\gamma \w Y \ts  \w Y$.\\
$-2\, \w G (V,\w Y)=0-2\gamma\la V,\w Y \ra +2\, e \la X_0, V \ra $ 
as $\la \w Y, \w Y\ra=1$ and $\la \w Y,X_0 \ra=0$.\\
So $$\Div \w F (V)=2\, e\la X_0, V \ra-(2\gamma+\w S)\la \w Y, V\ra$$ or by musical isomorphisms :
$$(\Div \w F)^\sharp =2\, e X_0-(2\gamma+\w S)\w Y.$$

\subsubsection{Free fall in an electrogravitational field.}
We want to prove that the vector field 
$$X:= X_0-\frac{e}{\mu}\w Y$$
is a geodesic vector field (where $\mu \neq 0$).

To prove this, one can just compute $\wnab_X X$, the only thing to notice being that \\
$X_0(\frac{e}{\mu})=0$. 

However, we prefer to show how we were lead to this simple but convincing result. As we said, a geodesic movement would have to imply the fifth dimension, and we want its projection on $H$ to be the flow of $X_0$. It was therefore natural to look for a vector field of the form $W=X_0+ \alpha \w Y$ for some function $\alpha$. Computing $\wnab_W W$, we get :
\begin{align*}
\wnab_{X_0+\alpha \w Y}(X_0+\alpha \w Y) & = \wnab_{X_0} X_0+(X_0(\alpha)+\alpha \w Y(\alpha) ).\w Y+\alpha (\wnab_{X_0} Y+\wnab_{\w Y} X_0)+\alpha ^2 \wnab_{\w Y} Y\\
                     & =\wnab_{X_0} X_0+(X_0(\alpha)+\alpha \w Y(\alpha) ).\w Y+\alpha (\wnab_{X_0} \w Y+\wnab_{\w Y} X_0)
\end{align*}
because $\w Y$ is geodesic. We now use a few facts : first, the connection is torsion-free, and $\w Y$ is killing. Therefore $[X_0,\w Y]=0$, and thus $\wnab_{X_0}\w Y=\wnab_{\w Y} X_0$. Secondly, $2\wnab_{X_0}\w Y=\, ^e\w F(X_0)$. And thirdly, $\wnab_{X_0}X_0=\frac{e}{\mu}. ^e\w F(X_0)$. So
$$\wnab_{X_0+\alpha \w Y}(X_0+\alpha \w Y)=(\frac{e}{\mu}+\alpha )\, ^e\w F (X_0)+ (X_0(\alpha)+\alpha \w Y (\alpha)).\w Y$$
But we have seen that $^e\w F(X_0)$ is horizontal, that is, orthogonal to $\w Y$. Therefore, $\wnab_W W$ will be zero if $\frac{e}{\mu}=-\alpha$ and $X_0(\alpha)+\alpha \w Y (\alpha)=0$. We thus have to check that 
$$X_0 (\frac{e}{\mu})-\frac{e}{\mu}.\w Y(\frac{e}{\mu})=0.$$
$\w Y$ generating isometries, $\w Y(\frac{e}{\mu})=0$. So it remains to check that $X_0 (\frac{e}{\mu})=0$.

This conservation law is a simple consequence of the two we already obtained : 
$$\Div (\mu X_0)=\Div (e X_0)=0.$$
 Indeed, developing these equations gives:
$$\mu .\Div X_0+X_0(\mu)=0$$
$$e.\Div X_0+X_0(e)=0$$
Multiplying the first equation by $e$ and the second by $\mu$, then subtracting,  gives $X_0 (\frac{e}{\mu})=0$.

\subsection{Remarks.}
\subsubsection{Remark on physical measurement in five dimensions.}
The equations obtained for a family of observers are general; they follow from the sole hypothesis of a fifth geodesic dimension. A fluid of matter is a choice of a geometric setting. Once a family of observers is defined, the equations obtained with these hypothesis can be seen as measures made by the observers. As such, and considering they can only "see" four dimensions, they can be made in two ways. Either these measures "neglect" everything happening on the fifth dimension, and this corresponds to projecting all the equations on $H$. Or the measures consists in taking the "mean" value along the fifth dimension, and this corresponds to going to the quotient, which is exactly the frame of Kaluza-Klein theory. In both cases, one recover the classical physical equations.

\subsubsection{The cosmological constant.}
If needed, one can easily introduce the cosmological constant $\Lambda$ in our model by considering a domain such that  $G+\Lambda g$ is a fluid as we defined it.

\subsubsection{Timelike fifth dimension and metric signature.}
We have chosen a spacelike fifth dimension as this is usually what is done. However, for the results of this paper, there is no need to do so. Indeed, choosing the fifth dimension to be timelike, that is choosing the restricition of the metric $\g$ to each fiber $S^1_x$ to be of timelike, introduces no change in the results and formulae obtained here. Essentially, one just need to replace the charge density $e$ by $-e$.

Definitions of fluid must be slightly modified to require that the 1-dimensional timelike eigenspace of $^eG$ should have a timelike $g$-orthonal projection on $Y^{\bot}$, (the fluid is not "flowing along the fifth dimension").

In fact, results obtained by Michel Vaugon, towards a geometrical frame for quantum mechanics in the spirit of this chapter, indicate that it might be useful, or even necessary, to consider such a signature for the metric. See the related paper indicated in section 7.5.

Note however that technical issues concerning Causality will appear. They can be ignored on the base space $\B \M$ by suitably taking "mean" values along the fifth dimension...

\subsubsection{Other geometrical-physical equations.}

The above equations were obtained using identities of Riemannian geometry . But there exist other such equations. One can then ask wether we could obtain other equations having a physical meaning.

For example, one can compute $trace_{\g} \w G$, which by definition is $-3/2\w S_{\g}$, but also 
$-\mu+\gamma$, from where we get another equation : $\mu-\gamma=3/2\w S_{\g}$.

These kind of equations already exist in classical 4-dimensional general relativity.

There is also another very important equation that wasn't used in this chapter, the Raychaudhuri equation seen in previous chapters (see Hawking-Ellis, Wald, Choquet-Bruhat). It gives the evolution of a family of geodesics defined as the flow lines of a vector field $X$ satisfying $\wnab_XX=0$. This equation, usually given in a 4-dimensional space-time, is valid in any dimension, and it gives important motion equations. It is a purely geometrical equation, lying on basic Riemannian geometry identities.

\section{Beyond five dimensions.}
Theoretical evidences, like string theory, suggest the need for a spacetime with more than five dimensions. We want to present in this section a possible extension of our model, that preserve the results obtained so far for the inclusion of electromagnetism, but that enable the possible inclusion of such other dimensions that might model geometrically other physical effects.
Although we do not pretend here to model precisely other known physical interactions, we present a geometric structure giving a possible way, for instance, to model or encode deviations from standard 4-dimensional General Relativity, or "dark" effects such as dark matter or energy.
 
\subsection{Multi-fiber bundle.}
\subsubsection{Multiple fibers at each point of spacetime.}
The mathematical translation of the heuristic idea of a 4-dimensional classical spacetime equipped with extra "small" dimensions is a fiber bundle structure $\pi : \M\rightarrow \B\M$ on a $(4+k)$-dimensional manifold $\M$, with fiber a compact manifold $\F$ of dimension $k$, more shortly a $\F$-fibration.

Now if we want to keep the result obtained for electromagnetism while including other possible interactions, the fiber $\F$ should be of the form $\F=S^1\times W$ where  $W$ is a compact manifold of dimension $m$ and  $S^1$ is the classical circle.  However, if we want to keep results obtained for electromagnetism in 5 dimensions, through objects naturally given by the action of $S^1$, we face a important issue : at each point $x\in \M$, such a fiber bundle gives a natural fiber $\F=S^1\times W$ through $x$, but it does not give a natural fiber through $x$ isomorphic to $S^1$ only ; there is no natural splitting of the fiber $S^1\times W$  at $x$. Therefore, such a fiber bundle alone will not furnish an electromagnetic potential $\w Y$. 

A very elegant extension of the structure of fiber bundle, giving a way to define any number of natural fibers at each point $x$ of a manifold $\M$, was originally proposed by Michel Vaugon in [ref]. We give here a new approach, based on the more classical notions of  fibrations and submersions.

\subsubsection{Splitting of a product manifold.}
A natural way to split a fiber $\F$ of the form $S\times W$ is based on the following nice construction. Let $\F$, $S$ and $W$ be three compact manifolds and let 
$$\Phi=(h,f) : \F\rightarrow S\times W$$
 be a diffeomorphism, where $h$ and $f$ are the components of $\Phi$. Then $h :\F\rightarrow S$ and $f : \F \rightarrow W$ are submersions. 
We then define unambiguously, for any $x\in\F$, two fibers at $x$ by :
$$ S_x:= f^{-1}(f(x))$$
$$ W_x:=h^{-1}(h(x))$$
Because $S$ and $W$ are compact, a theorem of Ehresmann states that these submersions are in fact fibrations.
Using the diffeomorphism $\Phi$, we can see that $h$ and $f$ are more precisely fibrations with fibers $W$ and $S$ respectively. Indeed, the restrictions 
$h|_{S_x} : S_x\rightarrow S$ and $f|_{W_x} : W_x\rightarrow W$ are diffeomorphisms whose inverse maps are given respectively by :
$$(h|_{S_x})^{-1}(u)=\Phi^{-1}(u,f(x)). \quad \quad  (f|_{W_x})^{-1}(v)=\Phi^{-1}(h(x),v).$$
Thus :
\begin{itemize}
\item $h :\F \rightarrow S$ is a $W$-fibration.
\item $f :\F \rightarrow W$ is a $S$-fibration.
\end{itemize}
Furthermore, we have a natural splitting of the manifold $\F$ as a product of two fibers at a given point : for a given point $p \in\F$, we have a natural diffeomorphism:
\begin{align}
\psi_p : \quad & \F \longrightarrow \quad   S_p \times W_p \nonumber \\
 & y \longmapsto \quad  (\, \Phi^{-1}(h(y),f(p))\, , \, \Phi^{-1}(h(p),f(y))\, ) \nonumber
\end{align}
Indeed, the inverse map is given by :
$$\psi_p^{-1}: S_p \times W_p \longrightarrow  \F$$
$$\quad \quad \quad \quad \quad \quad (a,b) \longmapsto \Phi^{-1}(h(a)\, , \, f(b))$$
To prove this, note first that, for $(a,b)\in S_p \times W_p$, by definition of the fibers, $f(a)=f(p)$ and $h(b)=h(p)$. Setting $y=\Phi^{-1}(h(a) , f(b))$, by definition of $\Phi$ and its components $h$ and $f$, $h(y)=h(a)$ and $f(y)=f(b)$. Therefore, $\Phi^{-1}(h(y), f(p))= \Phi^{-1}(h(a),  f(a))=a$ and $\Phi^{-1}(h(p), f(y))= \Phi^{-1}(h(b),  f(b))=b$.

\subsubsection{Multi-fiber bundle.}
We now apply this construction to define a \defin{multi-fiber} structure on a manifold $\M$ equipped with a classical fiber-bundle structure $\pi : \M\rightarrow \B\M$  whose fiber $\F=S\times W$ is a product of two compact manifolds. By definition, for any point $p\in \M$, there exist a \emph{bundle chart} $(\B \U, \phi)$ with $\pi(p)\in \B \U \subset \B\M$ :
\begin{align}
& \pi^{-1}(\B \U)   \stackrel{\phi}{\longrightarrow}\B \U\times \F =\B\U\times S \times W  \nonumber \\
\pi & \quad \downarrow \quad\quad  \swarrow p_1  \nonumber \\
& \quad \B \U \nonumber
\end{align}

Because of the commutativity of the diagram, $\phi$ can be written $\phi=(\pi,\Phi)$, with $\Phi: \pi^{-1}(\B \U) \rightarrow \F$ and $\Phi=(h,f)$, where $h : \pi^{-1}(\B \U) \rightarrow S$ and $f : \pi^{-1}(\B \U) \rightarrow W$ are the components of $\Phi$ as in the previous section. (To be rigorous, $\phi=(\pi |_{\pi^{-1}(\B\U )},\Phi)$. We also have here the following commutative diagram : 
\begin{align}
& \quad\pi^{-1}(\B \U)    \nonumber \\
(\pi,h)\swarrow \quad &  \quad \downarrow (\pi,\Phi) \quad\quad  \searrow (\pi,f)  \nonumber \\
\B\U \times S \quad\quad \longleftarrow \,\,\, &\B \U\times S\times W  \longrightarrow\quad\quad\B\U \times W \nonumber
\end{align}
where the horizontal arrows on the last line are the obvious projections.

We could therefore, using  $\Phi|_{\F_{ p}} : \F_{ p} \stackrel{\simeq}{ \rightarrow} S\times W$, where $\F_{ p}:=\pi^{-1}(\pi(p))$,  define fibers $S_p$ and $W_p$  :
 $ S_p := (f|_{\F_{p}})^{-1}(f(p))$ and $ W_p := (h|_{\F_{p}})^{-1}(h(p))$.

However, another chart $(\B \U', \phi')$ around $\B p:=\pi(p)$, with $\phi'=(\pi,\Phi')$ and $\Phi'=(h',f')$,  could give rise to different fibers $S_p$ and $W_p$ if for example $(f|_{\F_{p}})^{-1}(f(p))\neq(f'|_{\F_{p}})^{-1}(f'(p))$.

To get well-defined fibers $S$ and $W$ through $p$, we therefore need to impose a compatibility condition between the charts. This will lead to our definition of \defin{multi-fiber bundle}.

To get there, remember the following facts : as a fiber bundle, $\pi : \M\rightarrow \B\M$ can be considered as being equipped with a \emph{bundle atlas}, that is, a family $\{(\B\U_\alpha, \phi_\alpha)\}_{\alpha \in A}$ of bundle charts such that $\{\B\U_\alpha\}_{\alpha \in A}$ is a cover of $\B \M$. Then, if $\B\U_\alpha \cap \B\U_\beta$ is not empty, we have an \emph{overlap map} 
$$\phi_\alpha \circ\phi_\beta^{-1} : (\B\U_\alpha \cap \B\U_\beta)\times\F \rightarrow (\B\U_\alpha \cap \B\U_\beta)\times\F.$$
Also, still writing $\phi_\alpha=(\pi, \Phi_\alpha)$, $\Phi_\alpha |_{\F_p}$  is a diffeomorphism for each $p$ with $\pi(p)\in\B\U_\alpha$. Therefore 
$\Phi_\alpha |_{\F_p}\circ \Phi_\beta |_{\F_p}^{-1} :\F\rightarrow \F$ is a diffeomorphism for all $p$ such that $\pi (p) \in \B\U_\alpha \cap \B\U_\beta$.

It is on these overlap maps that we shall impose a compatibility condition : 
\begin{greybox}
Let $S$ and $W$ be two compact manifolds, and $\F:=S\times W$.

Let $\pi : \M\rightarrow \B\M$ be a fiber-bundle with fiber $\F$, and let $\{(\B\U_\alpha, \phi_\alpha)\}_{\alpha \in A'}$ be a complete bundle atlas for $\pi$. We have at each $p\in\M$ a global $\pi$-fiber : $\F_{ p}:=\pi^{-1}(\pi(p))$

Let's write $\phi_\gamma=(\pi,\Phi_\gamma)$ and $\Phi_\gamma=(h_\gamma, f_\gamma)$ for any bundle chart. We say that $\M$ is a \defin{multi-fiber bundle} with fibers $(S,W)$, or a $(S,W)$-fibration, if there exists a \emph{sub-atlas} $\{(\B\U_\alpha, \phi_\alpha)\}_{\alpha \in A}$ such that we have, for any $\alpha, \beta \in A$ with $\B\U_\alpha \cap \B\U_\beta\neq \emptyset$ and any $p\in \pi^{-1}(\B\U_\alpha \cap \B\U_\beta)$ :
\begin{itemize}
	\item $\phi_\alpha^{-1}(\{\pi(p)\}\times S\times \{f_\alpha(p)\})=\phi_\beta^{-1}(\{\pi(p)\}\times S\times \{f_\beta(p)\})$
	\item $\phi_\alpha^{-1}(\{\pi(p)\}\times\{h_\alpha(p)\}\times W)=\phi_\beta^{-1}(\{\pi(p)\}\times\{h_\beta(p)\}\times W)$
\end{itemize}
It is fair to call this sub-atlas $\{(\B\U_\alpha, \phi_\alpha)\}_{\alpha \in A}$ a multi-fiber atlas, and its charts, muti-fiber charts. With it, we can define unambiguously two new fibers at each point $p\in \M$ : using any chart $(\B\U_\alpha, \phi_\alpha)$ of this multi-fiber atlas with $\pi(p)\in\B\U_\alpha$ : 
\begin{itemize}
\item The $S$-fiber $ S_p := \phi_\alpha^{-1}(\{\pi(p)\}\times S\times \{f_\alpha(p)\})$, a submanifold of  $\F_{p}$
\item The $W$-fiber $\phi_\alpha^{-1}(\{\pi(p)\}\times\{h_\alpha(p)\}\times W)$, a submanifold of  $\F_{p}$
\end{itemize}
\end{greybox}

\emph{Notation : We shall  note $\B x=\pi(x)$ for $x$ in $\M$, and 
$\F_{\B x}:=(S \times W)_{\B x}:=\pi^{-1}(\pi(x)):= \pi^{-1} (\B x)$
the $\pi$-fiber at a point $x$ of $\M$. $(S \times W)_{\B x}$ is intuitively $\{\B x\}\times \F$. We also note $\B\U:=\pi(\U)$ for a set $\U$. For a map $f:\M\rightarrow \N$, where $\N$ is a manifold, we sometimes note $f_{\B x}$ the restriction of $f$ to the $\pi$-fiber $(S \times W)_{\B x}$ : $f_{\B x}:=f|_{\pi^{-1}(\B x)} :=f|_{\pi^{-1}(\pi(x))} :=f|_{\F_{\B p}}$.}

Note that the compatibility condition given in the definition of a multi-fiber bundle is equivalent to the following on the $\Phi_\alpha=(h_\alpha, f_\alpha)$'s :
\begin{greybox}
For any $p$ with $\pi (p) \in \B\U_\alpha \cap \B\U_\beta$, we have :
\begin{itemize}
	\item $\Phi_\alpha |_{\F_p}^{-1} (S\times \{f_\alpha(p)\})=\Phi_\beta |_{\F_p}^{-1} (S\times \{f_\beta(p)\})$
	\item $\Phi_\alpha |_{\F_p}^{-1} (\{h_\alpha(p)\}\times W)=\Phi_\beta |_{\F_p}^{-1} (\{h_\beta(p)\}\times W)$
\end{itemize}	
\end{greybox}

Using this and the splitting of each fiber $\F_{p}$ with the diffeomorphism $\Phi|_{\F_{ p}} : \F_{ p} \stackrel{\simeq}{ \rightarrow} S\times W$ of any multi-fiber chart, as seen in the previous section, we can give equivalent characterization of the fibers $S_p$ and $W_p$ for $p\in\M$ :

\begin{greybox}
\begin{itemize}
\item $ S_p := \Phi_\alpha |_{\F_p}^{-1} (S\times \{f_\alpha(p)\})=(f_\alpha|_{\F_{ p}})^{-1}(f_\alpha(p))$
\item $ W_p := \Phi_\alpha |_{\F_p}^{-1} (\{h_\alpha(p)\}\times W) = (h_\alpha|_{\F_{p}})^{-1}(h_\alpha(p))$
\end{itemize}
\end{greybox}

\begin{greybox}
The fundamental idea of multi-fiber structure is that, for a $(S\times W)$-fibration structure on a manifold $\M$, one can furthermore  define unambiguously, at each point, objects that depend only on one of the components, $S$ or $W$, of the global $(S\times W)$ fiber. See below.
\end{greybox}

The multi-fiber structure satisfies the following natural and important properties :
\begin{greybox}
\begin{itemize}
\item \emph{Fibers are well defined : } $p'\in S_p \Rightarrow S_{p'}=S_p$ and $p'\in W_p \Rightarrow W_{p'}=W_p$.
\item \emph{Splitting of the fibers : }Thanks to the splitting $\psi_p$ defined in the previous section,  we have at each point $p\in \M$ and for each multi-fiber chart $(\B\U_\alpha, (\pi,\Phi_\alpha))$ around $\B p$, a canonical isomorphism  :
$\psi_{p,\alpha} : \F_{p} \stackrel{\simeq}{ \rightarrow} S_p \times W_p$. The choice of another chart $(\B\U_\beta, (\pi,\Phi_\beta))$ around $\B p$ will lead to the same splitting but, via the overlap map, through diffeomorphisms of $S_p$ and $W_p$.
\item \emph{Adapted charts :} Let us fix a point $p\in\M$. A multi-fiber chart around $\B p$ gives a diffeomorphism 
$$\phi_{\U} :\U\rightarrow \B\U \times \F_{p}$$
 from a neighborhood $\U=\U_p$ of $p$ in $\M$. Composing with $Id\times \psi_p$, where $\psi_p$ is the splitting of the fiber  $ \F_{p}$ associated to $\Phi_\U$, we get an adapted diffeomorphism 
$$\varphi_{\U} =(Id\times \psi_p)\circ  \phi_{\U} : \U \rightarrow \B\U \times  S_p \times W_p$$
singularizing the fibers at $p$.  Then, taking coordinates charts on open subsets of  $ \B\U$, $S_p$ and $W_p$ respectively, we obtain very useful adapted charts on $\U_p$. See below.
\item \emph{Orientation of the fibers :} Let  $S$ be given an orientation ; an orientation on $W$ would be treated the same way. For any point $p\in\M$, the diffeomorphisms $h_{\alpha}|_{S_p} : S_p\rightarrow S$ can be used to pull-back the orientation of $S$ on $S_p$. We say that the multi-fiber structure  is compatible with the orientation of $S$ if, in some neighborhood $\U_p$ of any point $p$, there is a frame field for $\T_y S_y$, $y\in\U_p$, compatible with the orientation pulled-back by any $h_{\alpha}|_{S_y}$ such that $y\in\U_{\alpha}\cap\U_p$. 
\end{itemize}
\end{greybox}

We now consider $\M$ to be equipped with a metric $\g$. We then define the \defin{horizontal space} $H_p$ at a point $p\in\M$ as the $\g$-orthogonal space to $\T_p \F_{p}$ in $\T_p\M$ :
\begin{greybox}
$$H_p := (\T_p \F_{p})^\bot $$
\end{greybox}
We can then consider a compatibility condition between $\g$ and the multi-fiber structure:
\begin{greybox}
\begin{itemize}
\item \emph{Signature of the fibers :} We say that the metric  $\g$ is compatible with the multi-fiber bundle structure if the signature of the restriction of $\g$ to any fiber as defined above is constant. That is, for any $p \in M$, the signature of $\g$ restricted to $S_p$ and $W_p$ is independent of $p$. In this case, for given signatures $\sigma_a$ and $\sigma_b$ of adequate length, we will say that $\g$ is of signature $\sigma_a$ on $S$ and $\sigma_b$ on $W$. The signature of $\g_p$ on the horizontal space $H_p$ is then also independent of $p$.
\end{itemize}
\end{greybox}

Note that all the above construction can easily be generalized to define more than 2 fibers at each point of a manifold $\M$. If we are given a global $\pi$-fiber of the form $\F=W_1\times ... \times W_k$, we essentially replace the compatibility conditions on the charts by something like :  
$$\phi_\alpha^{-1}(\{\pi(p)\}\times\{f^1_\alpha(p)\}\times...\times W_j\times...\times \{f^k_\alpha(p)\})=\phi_\beta^{-1}(\{\pi(p)\}\times\{f^1_\beta(p)\}\times...\times W_j\times...\times \{f^k_\beta\})$$
with adapted analog properties. 
\\

\footnotesize{Remark : A simple idea to define a multi-fiber structure based on the splitting of the fibers could have been to consider on the initial $\F$-fibration $\pi : \M\rightarrow \B\M$, with $\F=S\times W$, an additional map $\Phi=(h,f) :\M\rightarrow S\times W$ such that, for any $p\in\M$, the restriction 
$\Phi|_{\F_{\B p}} : \F_{\B p} \stackrel{\simeq}{ \rightarrow} S\times W$
 of $\Phi$ to the $\pi$-fiber $\F_{\B p}:=\pi^{-1}(\pi(p))$ is a diffeomorphism ;  hence the following diagram :
 \begin{align}
& \M  \stackrel{\Phi}{\longrightarrow}S \times W \nonumber \\
\pi & \downarrow \nonumber \\
& \B\M \nonumber
\end{align}
Then, simply define $ S_p := (f|_{\F_{\B p}})^{-1}(f(p))$ and $ W_p := (h|_{\F_{\B p}})^{-1}(h(p))$. But it is easy to see that in fact this gives a trivial fibration in the sense that $\M$ is then diffeomorphic to $\B\M \times S \times W$ : just consider $\phi : \M \rightarrow \B\M \times S \times W$, $x \mapsto (\pi (x), (h(x),f(x)))=(\pi(x),\Phi(x))$ whose inverse is 
$\phi^{-1} : (a,b)\mapsto (\Phi|_{\pi^{-1}(a)})^{-1}(b)$.
}
\normalsize

\subsubsection{Adapted charts.} 
As they are useful  to understand the situation, let us see how we get adapted charts, and what they look like. We take here $S=S^1$ as we will be mostly interested in this case. Consider $\M$ with a $(S^1\times W)$-multi fiber structure. As we saw above, starting with a trivialization chart of the fibration $\pi$ and composing with the splitting of the fiber  $ \F_{p}$ by $\psi_p$, we have in the neighborhood of any fixed point $p\in\M$ an adapted diffeomorphism of the following form:
\begin{align}
 \U_p  \stackrel{\phi_\U}{\longrightarrow} \, \, & \B\U_p \times (S^1\times W)_p \nonumber \\
& Id \downarrow \,\,\, \, \quad \quad \,\, \downarrow \psi_p  \nonumber \\
& \B\U_p \times \, S^1_p\times \, W_p \nonumber
\end{align}
As $S^1_p$ and $W_p$ are diffeomorphic to $S^1$ and $W$ respectively, we can now take coordinates $(x^i)$ on $\B\U_p$, $(u)$ on some neighborhood $\dot S^1_p$ of $p$ in $S^1_p$, and $(w^k)$ on some neighborhood $\W_p$ of $p$ in $W_p$, to obtain a chart of the form :
\begin{align}
 \U_p  \stackrel{\phi_\U}{\longrightarrow} \, \, & \B\U_p \times (S^1\times W)_p \nonumber \\
& Id \downarrow \,\,\, \, \quad \quad \,\, \downarrow \psi_p \nonumber \\
& \B\U_p \times \, \dot S^1_p\times \, \W_p \nonumber \\
& \downarrow \quad \quad \downarrow \quad \quad \downarrow   \nonumber \\
& (x^i, \quad  \, u, \,\quad \, w^k) \nonumber
\end{align}
Centering the chart so that the coordinates of $p$ are $(0,...,0)$, the coordinates expression of $f|_{\F_{p}}$ and $h|_{\F_{p}}$ are :
$$f|_{\F_{p}}(0,...,0, u,w^1,...,w^m)=(w^1,...,w^m)$$
$$h|_{\F_{p}}(0,...,0, u,w^1,...,w^m)=(u)$$
Indeed, they are submersions ! It  is then clear that :
$$(f|_{\F_{p}})^{-1}(f(x))=\{(0,..0,u,0,...,0)\}:=\dot S^1_p$$
$$(h|_{\F_{p}})^{-1}(h(x))=\{(0,...,0,0,w^1,...,w^m)\}:=\W_p$$
for some neighborhoods $\dot{S^1_p}$ and $\W_p$ of $p$ in $S^1_p$ and $W_p$ respectively, where the coordinates are defined. 

\subsubsection{Multi-fiber manifolds.} 
Our definition of the multi-fiber bundle structure is based on the fiber bundle atlas. We can also define directly on a manifold a mean to get well-defined fibers at each point, without referring to an existing bundle structure. This was the original idea of the second author, Michel Vaugon.
\begin{greybox}
Let $\M$ be a differential $n$-dimensional manifold, and let $S$ and $W$ be two compact manifolds of respective dimension $k$ and $l$.
A diffeomorphism : $$\phi:\U \rightarrow \B\U\times S \times W$$ where $\U$ is an open set in $\M$ and $\B\U$ an open set in  $\R^{n-k-l}$, will be called an \defin{observation diffeomorphism}, and the couple $(\U,\phi)$ an \defin{observation chart} (if $\B\U$ is understood). We note $\phi=(f^1,f^2,f^3)$ the threee components of $\phi$.

We say that $\M$ is a \defin{multi-fiber manifold} with fibers $S$ and $W$ if there exists a \defin{$(S,W)$-observation atlas}, that is, a family $\{(\U_\alpha, \phi_\alpha)\}_{\alpha \in A}$ of observation charts such that $\cup\U_\alpha=\M$, and satisfying for any $\alpha, \beta \in A$ and any $p\in \U_\alpha \cap \U_\beta$  :
\begin{itemize}
	\item $\phi_\alpha^{-1}(\{ f^1_\alpha(p)\}\times S\times \{f^3_\alpha(p)\})=\phi_\beta^{-1}(\{f^1_\beta(p)\}\times S\times \{f^3_\beta(p)\})$
	\item $\phi_\alpha^{-1}(\{f^1_\alpha(p)\}\times\{f^2_\alpha (p)\}\times W)=\phi_\beta^{-1}(\{f^1_\beta(p)\}\times\{f^2_\beta(p)\}\times W)$
\end{itemize}
The definitions, for any $p\in \M$, of fibers $S_p$ and $W_p$ is then given as above.	
\end{greybox}
An observation atlas can be completed in a \emph{complete observation atlas} in the same manner as for a classical differential manifold atlas.

Whereas the multi-fiber bundle structure is a generalization of the fiber bundle structure (or fibration), the multi-fiber manifold structure can be seen as a generalization of the \emph{foliation} structure.

\emph{Remark and proposition : if we consider a single compact fiber W, and an observation atlas of observation charts of the form $\phi:\U \rightarrow \B\U\times W$ satisfying the adapted compatibility condition : $\phi_\alpha^{-1}(\{f^1_\alpha(p)\}\times W)=\phi_\beta^{-1}(\{f^1_\beta(p)\}\times W)$,
it is fairly easy to prove that $\M$ can be equipped with a fiber bundle structure $\pi : \M\rightarrow \B\M$ with fiber $W$ for some manifold $\B\M$. Indeed, consider $\sim$ defined by $p\sim p'$ if $\phi_\alpha^{-1}(\{f^1_\alpha(p)\}\times W)=\phi_\alpha^{-1}(\{f^1_\alpha(p')\}\times W)$, that is, if $W_p=W_{p'}$. Define $\B\M:=\M/\sim$. The compacity of $W$ ensures $\B\M$ is Hausdorf ; if $W$ is not compact, one must add the requirement that $\phi_\alpha^{-1}(\{f^1_\alpha(p)\}\times W)$ is closed in $\M$ for all $\alpha$ and all $p$.}
\\

The main difference between this structure and the multi-fiber bundle structure, is that it makes no reference to a "natural" manifold $\B\M$ linked to the "horizontal" distribution $H_p=(S_p \times W_p)^\bot$. The use of the bundle structure can bring formulas \emph{à la O'Neill} linking the geometries of $\B\M$, $S$ and $W$. On a heuristic point of view, keeping the bundle structure keep the idea of "small compact dimensions" attached to classical 4-dimensional spacetime $\B\M$, whereas choosing as model a multi-fiber manifold with a $(S,W)$-observation atlas is  more radical as it makes no reference to a specific 4-dimensional manifold.

\subsubsection{Construction.}
We imitate here the classical constructions of fiber bundles using cocyles with value in a subgroup of the diffeomorphisms group of the fiber. This starts by observing that the overlap maps $\phi_\alpha \circ\phi_\beta^{-1} : (\B\U_\alpha \cap \B\U_\beta)\times\F \rightarrow (\B\U_\alpha \cap \B\U_\beta)\times\F$ give rise to diffeomorphisms of the fiber $\F$ : $$\Phi_{\alpha\beta}(p):=\Phi_\alpha |_{\F_p}\circ \Phi_\beta |_{\F_p}^{-1} :\F\rightarrow \F, $$ where as always, $\phi$ is written $\phi=(\pi,\Phi)$.
In our multi-fiber case, with $\F=S\times W$, we have two more diffeomorphisms, writing again $\Phi_\alpha=(h_\alpha, f_\alpha)$ :
$$h_{\alpha\beta}(p):=h_\alpha |_{S_p}\circ h_\beta |_{S_p}^{-1} :S\rightarrow S $$
$$f_{\alpha\beta}(p):=f_\alpha |_{W_p}\circ f_\beta |_{W_p}^{-1} :W\rightarrow W. $$
Indeed, $h_\alpha |_{S_p}:S_p\rightarrow S$ and $f_\alpha |_{W_p}:W_p\rightarrow W$ are diffeormorphisms. We therefore have, for each $\alpha,\beta$ such that $\B\U_\alpha \cap \B\U_\beta\neq \emptyset$,  maps $p\mapsto h_{\alpha\beta}(p) \in Diff(S)$ and $p\mapsto f_{\alpha\beta}(p) \in Diff(W)$.
These diffeomorphisms satisfy a cocycle relation, in the sense that : 
$$h_{\alpha\alpha}(p)=Id_S\,\,,\,\, h_{\alpha\beta}(p)=h_{\beta\alpha}(p)^{-1}\,\,,\,\, h_{\alpha\beta}(p)\circ h_{\beta\gamma}(p)=h_{\alpha\gamma}(p),$$ and similarly for $f_{\alpha\beta}(p)$. 
These cocycles are the building blocks for our multi-fiber structure.\\

So let $\B\M, S,W$ be three manifolds, $S$ and $W$ being compact. Let $H$ and $F$ be two Lie groups that acts on the left on $S$ and $W$ respectively. Let be given an open cover $(\B\U_\alpha)_{\alpha\in A}$ of $\B\M$. A $H$-cocycle for $(\B\U_\alpha)$ is the assignment of a smooth map $h_{\alpha\beta}:(\B\U_\alpha \cap \B\U_\beta)\rightarrow H$ to every nonempty intersection $\B\U_\alpha \cap \B\U_\beta$ such that the cocycle conditions holds for $h_{\alpha\beta}$. We suppose we are given an $H$-cocycle $(h_{\alpha\beta})$ and a $F$-cocycle $(f_{\alpha\beta})$, and from now on, to make the writings simpler, we do as if $H$ and $F$ were subgroups of the diffeomorphisms groups $Diff(S)$ and $Diff(W)$ respectively, with the natural actions ; the reader will easily adapt what follows to the more general case of Lie groups acting on $Diff$.

With these datas, we can construct our $(S,W)$-multi-fiber strucure over $\B\M$. For this, we consider the disjoint union
$$\Sigma:=\coprod_\alpha\, \{\alpha\}\times\B\U_\alpha\times S \times W$$
and on $\Sigma$, the equivalence relation $\sim$ defined by $$(\alpha, p, x,y)\sim (\beta, p',x',y')\quad \mbox{iff}$$ $$p=p',\quad x'=h_{\alpha\beta}.x,\quad \mbox{and} \quad y'=f_{\alpha\beta}.y$$
We define $\M:=\B\M/\sim$ and $\pi :\M\rightarrow\B\M, \, [\alpha, p,x,y]\mapsto p$, where $ [\alpha, p,x,y]$ is the equivalence class of $(\alpha, p,x,y)$. 
Now, defining $\Phi_{\alpha\beta}:=(h_{\alpha\beta},f_{\alpha\beta}):\F\rightarrow\F$ for $\F:S\times W$, we obtain a $H\times F$-cocycle acting on $\F$. Thanks to this, it is classical to prove that $\M$ has a $\F$-bundle structure.

Indeed, an bundle atlas on $\M$ is obtained by defining $\U_\alpha=\pi^{-1}(\B\U_\alpha)$ and $\phi_\alpha :\U_\alpha\rightarrow\B\U_\alpha\times S \times W$ as the map $m\in \U_\alpha\mapsto(p,x,y)$ such that $(\alpha,p,x,y)\in m$ ; remember that $m$ is an equivalence class. This is well defined because, as $h_{\alpha\alpha}=Id_S$ and $f_{\alpha\alpha}=Id_W$, $(\alpha,p,x,y)\sim(\alpha,p',x',y')$ iff $p=p'$, $x=x'$ and $y=y'$. The family $(\B\U_\alpha, \phi_\alpha)_{\alpha\in A}$ then constitutes a $\F$-bundle atlas.

We won't go in the detailed proof of this. We just indicate what an overlap map $\phi_\alpha \circ \phi_\beta^{-1}$
looks like. So let $\B\U_\alpha\cap\B\U_\beta\neq\emptyset$, and $p\in \B\U_\alpha\cap\B\U_\beta$. $\phi_\beta^{-1}(p,x,y)=[\beta,p,x,y]\in\pi^{-1}(\B\U_\alpha)$ thus $[\beta,p,x,y]=[\alpha,q,x',y']$ which means $p=q$, $x'=h_{\alpha\beta}.x$ and $y'=f_{\alpha\beta}.y$ ; i.e. $\phi_\beta^{-1}(p,x,y)=[\alpha,p, h_{\alpha\beta}.x, f_{\alpha\beta}.y]$. Therefore $\phi_\alpha \circ \phi_\beta^{-1}(p,x,y)=(p, h_{\alpha\beta}.x, f_{\alpha\beta}.y)=(p,\Phi_{\alpha\beta}(x,y))$. Setting a smooth structure on $\M$ and proving that it is Hausdorf is classic.
\\

Remark : A general diffeomorphism $\Phi=(h,f) : S\times W \rightarrow S\times W$ is written $\Phi(x,y)=(h(x,y),f(x,y))$. Here, the cocycle $\Phi_{\alpha\beta}$ are of the special form : $\Phi_{\alpha\beta}(p)(x,y)= (h_{\alpha\beta}(p)(x),f_{\alpha\beta}(p)(y))$. \\

Now, we need to check that we can extract from the $\F$-bundle atlas $(\B\U_\alpha, \phi_\alpha)_{\alpha\in A}$ a $(S,W)$-multi-fibers atlas. Once again, we write, $\phi_\alpha=(\pi,\Phi_\alpha)=(\pi,h_\alpha,f_\alpha)$.
  We have to prove that for any $m\in \pi^{-1}(\B\U_\alpha \cap \B\U_\beta)$ we have
$\phi_\alpha^{-1}(\{\pi(m)\}\times S\times \{f_\alpha(m)\})=\phi_\beta^{-1}(\{\pi(m)\}\times S\times \{f_\beta(m)\})$ and similarly for $W$. 

If $m\in \pi^{-1}(\B\U_\alpha \cap \B\U_\beta)$, it can be written $m=[\alpha, p,x,y]=[\beta, p,x',y']$ with $p=\pi(m)$, $x'=h_{\alpha\beta}.x$, $y'=f_{\alpha\beta}.y$. If $u\in \phi_\alpha^{-1}(\{\pi(m)\}\times S\times \{f_\alpha(m)\})$, $u$ is written $u=[\alpha,p',a,b]$. But $\pi(u)=\pi(m)$ implies $p=p'$ by definition of $\pi$. Similarly, by definition of $\phi_\alpha$, $\phi_\alpha(u)=(p,a,b)$ and $\phi_\alpha(m)=(p,x,y)$, so $f_\alpha(u)=f_\alpha(m)$ implies $b=y$. Therefore $u=[\alpha, p, a, y]$. Now, as $\pi(u)=\pi(m)$, $u\in \pi^{-1}(\B\U_\alpha \cap \B\U_\beta)$. So $u=[\beta, p, a',b']$ with $a'=h_{\alpha\beta}.a$ and $b'=f_{\alpha\beta}.y$. By definition of $\phi_\beta$, $\phi_\beta (u)=(p,a',b')$. But $a'=h_{\alpha\beta}.a\in S$ and $b'=f_{\alpha\beta}.y=y'=f_\beta(m)$. Therefore, $\phi_\beta(u)\in \{\pi(m)\}\times S\times \{f_\beta(m)\}$.

We just proved that $\phi_\beta^{-1}(\{\pi(m)\}\times S\times \{f_\beta(m)\})\subset \phi_\alpha^{-1}(\{\pi(m)\}\times S\times \{f_\alpha(m)\})$. Exchanging the role of $\alpha$ and $\beta$ above, we get that $\phi_\alpha^{-1}(\{\pi(m)\}\times S\times \{f_\alpha(m)\})=\phi_\beta^{-1}(\{\pi(m)\}\times S\times \{f_\beta(m)\})$. We proceed analogously to prove that $\phi_\alpha^{-1}(\{\pi(m)\}\times\{h_\alpha(m)\}\times W)=\phi_\beta^{-1}(\{\pi(m)\}\times\{h_\beta(m)\}\times W)$.

Therefore, $(\B\U_\alpha, \phi_\alpha)_{\alpha\in A}$ is also a $(S,W)$-multi-fiber bundle atlas. (Note however that $(\B\U_\alpha, \phi_\alpha)_{\alpha\in A}$ might be completed in a larger $ç(S\times W)$-bundle atlas).

\subsubsection{Building objects on fibers.}
Conversly, let $(\B\U_\alpha, \phi_\alpha)_{\alpha\in A}$ be a $(S,W)$-multi-fiber bundle atlas for a manifold $\M$, coming from a $\F$-bundle atlas for the fibration $\pi : \M \rightarrow\B\M$ where $\F=S\times W$. For each $p\in \M$, the fibers $S_p$ and $W_p$ are well defined. Again, we write $\phi_\alpha=(\pi,\Phi_\alpha)=(\pi,h_\alpha,f_\alpha)\,$. Then, 
for any $\alpha,\beta$ such that $\pi(p)\in \B\U_\alpha\cap\B\U_\beta$, $h_\alpha|_{S_p}$ and $h_\beta|_{S_p}$ are diffeomorphisms $S_p\rightarrow S$, and $f_\alpha|_{S_p}$ and $f_\beta|_{S_p}$ are diffeomorphisms $W_p\rightarrow W$.

For any $\alpha,\beta$ with $\B\U_\alpha\cap\B\U_\beta\neq \emptyset$ we therefore have maps :
$$h_{\alpha\beta} : p\mapsto h_{\alpha\beta}(p):=h_\alpha|_{S_p}\circ h_\beta|_{S_p}^{-1}\in Diff(S)$$
and
$$f_{\alpha\beta} : p\mapsto f_{\alpha\beta}(p):=f_\alpha|_{W_p}\circ f_\beta|_{W_p}^{-1}\in Diff(W).$$
For each $p\in\M$, $H_p:=(h_{\alpha\beta}(p))$ and $F_p:=(f_{\alpha\beta}(p))$ define cocycles in $Diff(S)$ and $Diff(W)$ respectively. We consider the subgroups $H$ of $Diff(S)$ and $F$ of $Diff(W)$ generated by the $H_p$ and $F_p$ respectively : $H:=<(\cup_{p\in\M} H_p>)$ and $F:=<(\cup_{p\in\M}F_p)>$.

We suggest here some possible ways to build tensors on $\M$ out of similar objects defined on $S$ or $W$, that could not be defined using the sole $\F$-bundle structure of $\M$. (We will use the fiber $S$, but obviously the same constructions can be made using $W$). \\

\defin{Functions :} Let $f\in\C^\infty(S)$. Let's say $f$ is $H$-invariant on $S$ if $\forall h\in H$, $h^*f=f$, that is, $f\circ h=f$. We can then define a smooth function $\tilde f$ on $\M$ by setting $\tilde f(p)=f(h_\alpha(p))$ for any multi-fiber chart $(\B\U_\alpha, \phi_\alpha)$, such that $p\in\U_\alpha:=\pi^{-1}(\B\U_\alpha)$.

The idea if of course that if $p\in\U_\beta$, because $h_\alpha|_{S_p}\circ h_\beta|_{S_p}^{-1}=h_{\alpha\beta}(p)\in H$, we have $f(h_\beta(p))=f((h_\alpha|_{S_p}\circ h_\beta|_{S_p}^{-1})(h_\beta(p))=f(h_\alpha(p))$.

We could not build such a function on $\M$ from a function on $S$ with the sole $\F$-bundle structure.
\\

\defin{Vector fields :} Let $X\in \Gamma (TS)$ be a vector field on $S$. Let's say $X$ is $H$-invariant on $S$ if $\forall h\in H$ $h_* X=X$, that is, $\forall x\in S$, $T_x h(X(x))=X(h(x))$. 

Let $p\in\M$ with $p\in\U_\alpha$ for some $(\B\U_\alpha, \phi_\alpha)$ of the multi-fiber atlas. Then $h_\alpha|_{S_p}:S_p\rightarrow S$ is a diffeomorphism. We define $\tilde X(p)= (T_p h_\alpha|_{S_p})^{-1}(X(h_\alpha(p))$.

If $p\in\U_\beta$, we have, as $h_\alpha|_{S_p}\circ h_\beta|_{S_p}^{-1}=h_{\alpha\beta}(p)\in H$ :
\begin{align}
	(T_p h_\beta|_{S_p})^{-1}(X(h_\beta(p))&=(T_p h_\beta|_{S_p})^{-1}(X(h_\beta|_{S_p}\circ h_\alpha|_{S_p}^{-1}(h_\alpha(p))\\
	&=(T_p h_\beta|_{S_p})^{-1}T_{h_\alpha(p)}(h_\beta|_{S_p}\circ h_\alpha|_{S_p}^{-1})(X(h_\alpha(p))\\
	&=(T_p h_\alpha|_{S_p})^{-1}(X(h_\alpha(p))
\end{align}
So $\tilde X$ is a well-defined vector field on $\M$ that, here again, could not have been defined from a vector field $X$ on $S$ without the multi-fiber structure.
\\

\defin{The case of covariant tensors :} Pulling back (covariant) tensors on $\M$ from tensors on $S$ is more delicate, as $h_\alpha$ only induces an isomorphism on the subspace $T_pS_p$ of the whole tangent space $T_p\M$. So even though we can push forward a vector $v$ of $T_p\M$ using $T_p h_\alpha$, $T_{h_\alpha(p)}(h_\beta|_{S_p}\circ h_\alpha|_{S_p}^{-1})(T_p h_\alpha(v))$ will not be well-defined outside $T_pS_p$.

It appears that the only way to define properly a tensor $\tilde A$ on $\M$ from the tensor $A$ on $S$ is to suppose that we are given an \emph{horizontal distribution} $\Hor$ on $\M$, that is, a smooth family $(H_p)_{p\in\M}$ of subspaces of $T_p\M$, such that at each $p\in\M$, $T_p\M=H_p\oplus T_p\F$ ; in this case , we also have $\T_p\M=H_p\oplus T_pS\oplus T_pW$.

For example, if $\M$ comes equipped with a Riemannian metric $\g$, an obvious choice is to take $H_p:=( T_pS\oplus T_pW)^\bot$. If $\g$ is semi-Riemannian, we have to require appropriate signature compatibility on the fibers to ensure that $H_p$ so defined is a supplementary to $ T_pS\oplus T_pW$ in $\T_p\M$.

We therefore suppose now that we are given such an horizontal distribution $\Hor$. Any tangent vector $v\in T_p\M$ can be written uniquely $v=v_H+v_S+v_W$, with $v_H\in H_p$, $v_S\in T_pS_p$ and $v_W\in T_pW_p$.

So let's $A\in T^{2,0}(S)$ be a covariant 2-tensor field on $S$. 
We shall say that $A$ is $H$-invariant on $S$ if, $\forall h\in H$, $h^*A=A$, that is : $A_{h(x)}(T_xh.u,T_xh.v)=A_x(u,v)$, $\forall x\in S$, $\forall u,v \in T_xS$.

Let again $p\in\M$ with $p\in\U_\alpha$ for some $(\B\U_\alpha, \phi_\alpha)$ in the multi-fiber atlas, and let $u,v\in T_p\M$. We define a (2,0)-tensor field $\tilde A$ on $\M$ by setting :
$$\tilde A_p (u,v)=((h_\alpha|_{S_p})^*A)_p(u_S,v_S):=A_{h_\alpha(p)}(T_p h_\alpha|_{S_p})(u_S),T_p h_\alpha|_{S_p})(v_S))$$
We check, as above in the case of a vector field, that $\tilde A$ is a well-defined (2,0)-tensor field on $\M$.

 \subsubsection{Building Metrics.}
 A natural question when given a classical fiber bundle $\pi:\M\rightarrow B$ over a manifold base $B$ and fiber $\F$, and metrics $\g_B$ and $\g_\F$ on $B$ and $\F$ respectively, is to build a natural metric $\g$ on $\M$ out of $\g_B$ and $\g_\F$.

However, just as we saw above for any covariant tensors, the difficulty to pull back tensors from $B$ or $\F$ is due to the absence of a canonical supplementary space to $T_p \F$ in $T_p\M$. 

As this natural question extends naturally to our multi-fiber structure case, and as a solution lies on the same requirement (which is the existence of a given horizontal distribution), we address it here in this more general case.

So let $\pi : \M\rightarrow\B\M$ be a $(S,W)$-multi-fiber bundle, and let be given the same cocycles and diffeomorphisms groups datas as at the beginning of the previous section.

We suppose that $\B\M$ is equipped with a metric $\g_B$ ($B$ for Base !), that $S$ is equipped with a $H$-invariant metric $\g_S$, and $W$ with a $F$-invariant metric $\g_W$. (Invariance here is understood as in the above case of covariant tensors in the previous section).
Define the vertical space $\V_p:=Ker (T_p\pi)$, let $\V=\cup_p \V_p$ be the \emph{vertical distribution}. We suppose we are given, at each $p\in\M$, a supplementary space $H_p$ to $\V_p$, called \emph{horizontal space}, and that the distribution $\Hor:=\cup_p H_p$ is a smooth distribution. At each $p\in\M$, we have :
$$T_p\M=\V_p\oplus H_p$$
$$\V_p=T_p\F_p=T_pS_p\oplus T_pW_p$$
And therefore : $T_p\M=H_P\oplus T_pS_p\oplus T_pW_p$.

Any $v\in T_p\M$ can thus be written : $v=v_H+v_S+v_W$, with $v_H\in H_p$, $v_S\in T_pS_p$ and $v_W\in T_pW_p$.

It is now easy to define a metric $\g$ on $\M$. Letting $p$ be any point in $\M$ with $p\in\U_\alpha$ for some $(\B\U_\alpha, \phi_\alpha)$ in the multi-fiber atlas, and $u,v\in\T_p\M$, define :
$$\g_p(u,v):=\pi^*\g_B(u_H,v_H)+(h_\alpha|_{S_p})^*\g_S(u_S,v_S)+(f_\alpha|_{W_p})^*\g_W(u_W,v_W).$$

Remember our notation from the previous section : if $\w t$ is a covariant tensor on $S$ or $W$, $\w {\tilde t}$ is the tensor pulled back by $h_\alpha|_{S_p}$ or $f_\alpha|_{W_p}$ respectively. We can also write the above definition of $\g$ :
$$\g:=\pi^*\g_B+\w{\tilde \g}_S+\w{\tilde \g}_W.$$
It is also now very easy to imagine \emph{warped metrics} in the spirit of O'Neill warped product : letting $a,b\in\C^\infty(\B\M)$ be positive functions, set :
$$\g:=\pi^*\g_B+(a\circ\pi)^2.\w{\tilde \g}_S+(b\circ\pi)^2.\w{\tilde \g}_W.$$

O'Neill type formulae and results can then be obtained linking geodesics or Ricci curvature (for example) of $(\M,\g)$ to the geometries of $(\B\M,\g_B)$, $(S,\g_S)$ and $(W,\g_W)$.

More generally, in [19], Michel Vaugon uses a metric conformal to the above, where the conformal factor function, $f\in\C^\infty(\M)$, $f>0$, is used to model quantum phenomena :
$$\g:=f^2.\pi^*\g_B+(a\circ\pi)^2.\w{\tilde \g}_S+(b\circ\pi)^2.\w{\tilde \g}_W.$$

\subsubsection{The special case of $S^1$, electromagnetic potential, and an example :} 
For our $(S,W)$-multi-fiber bundle structure, the case $S=S^1$, $S^1$ being the standard circle,  will be particularly important. Indeed, it is the fiber to be used to include electromagnetism in the geometric frame of general relativity according to Kaluza-Klien idea.

Besides, it is a very tractable case, as in particular we can define easily on $\M$ a natural vector field $\w Y$ associated to each fiber $S_p$ without using all the machinery of cocycles and invariance ; $\w Y$ will of course be the electromagnetic potential.

Indeed, consider $\M$, a $(S^1,W)$-multi fiber bundle. A circle fiber $S^1_p$ is well defined at each $p\in\M$. We suppose the multi-fiber structure on $\M$ is compatible with the natural orientation of $S^1$ (considering for example $S^1$ as $\R / \Z$). Then we can define unambiguously $\w Y$ to be the vector field defined at each $x \in \M$ to be tangent to the fiber $S^1_x$ and such that $\g (\w Y,\w Y)=-1$, with the chosen orientation for $S^1$. \\

As we already said, the main idea leading to the multi-fiber structure is that we will use the other fibers $W_p$ to model other physical interactions, but still keeping the possibility to use the geometry of the total fiber $\F_p \simeq S^1_p \times W$.

As an example we cite the use by the second author, Michel Vaugon, of a 
\defin{$(S^1\times S^3)$-multi-fiber structure on a manifold $\M$.}

The 3-dimensional sphere $S^3$ is the classical geometric space used in quantum theory to describe \emph{spin}. Indeed, $S^3$ carries a natural frame field as well as canonical endomorphisms giving precisely the spin matrixes. This lead Michel to use a multi-fiber bundle structure on a manifold $\M$ with fiber $\F=S^1\times S^3$ to give a geometric model of electromagnetic \emph{and} spin effects on $\M$. The $S^1$ component was used to define the electromagnetic potential, and the $S^3$ component to define objects related to the spin. Furthermore, the modelization of the physical effects of both electromagnetism and spin, for instance to describe the Stern and Gerlach experiment, required to use the geometry of the \emph{total} fiber  $\F=S^1\times S^3$, so the full multi-fiber structure was used. 

More precisely, this $S^1\times S^3$-multi-fibers structure, and the objects described below that can be built with it,  was the geometric starting point for the second author to build a new approach towards a geometric unification of General Relativity and Quantum Physics. It is based on special metrics on a $(5+k)$-dimensional manifold modelizing quantum particles physics in spacetime. In this setting, only the metric is relevant, no objects or laws are added, these appear as geometric quantities issued from curvature and geometric theorems, such as Bianchi identity, linked with the objects described below. See [19].

We therefore consider that the spacetime $(\M,\g)$ is a $(S^1,S^3)$-multi-fibers bundle.
At each point $x\in\M$, we then have two naturally defined fibers : $S^1_x$ and $S^3_x$.

As before, $S^1$ gives the electromagnetic potential $\w Y$ :  $\w Y$ is the vector field defined at each $x \in \M$ to be tangent to the fiber $S^1_x$ and such that $g (Y,Y)=-1$, with the chosen orientation for $S^1$.

Then, on $S^3$, any function, vector field, or covariant tensor, invariant by a subgroup of $Diff(S^3)$ containing the cocycles induced by the overlap maps of the multi-bundle atlas can be used to construct analog objects on $\M$. 
In his work for example, Michel uses the spectral theory of $S^3$ to transpose a Hilbertian basis of $L^2$ functions from $S^3$ to $\M$.

These objects could not be defined with a simple $(S^1\times S^3)$-fiber bundle structure on $\M$.

\subsection{Model for spacetime.}
We now propose our final model for space-time. For upcoming aesthetic reasons, and further developments by Michel Vaugon, we choose the fifth dimension to be timelike. The reader uncomfortable with the two timelike dimensions, can still consider the signature on $S^1$ to be spacelike; only minor sign changes will be required in front of expressions using $e$ or $Y$, but all what follows remains in fact essentially unchanged.
\begin{greybox} 
Let $S^1$ be the classical circle with a chosen orientation, and let $W$ be a compact manifold of dimension $m$. 

A \emph{Spacetime} is a semi-Riemannian manifold $(\M,\g)$ of dimension $5+m$ equipped with a multi-fiber bundle structure, of fiber $\F=S^1\times W$ as defined above. 

We suppose that the metric  $\g$ is compatible with the multi-fiber bundle structure, $\g$ being of  signature $(-1)$ on $S^1$, and $(+,...,+)$ on $W$ ; the total signature of $\g$ is thus $(-,+,+,+,-,+,...,+)$. $\B\M$ is "classical" spacetime.

The \defin{horizontal space} at a point $x$ is $H_x := (\T_x(S^1\times W)_{x})^\bot $, and $\g_x$ is of signature $(-,+,+,+)$ on $H_x$. $H_x$ represents the local and classical Minkowski spacetime at $x$.
\end{greybox}

The effect of the "extra" $m$ dimensions carried by $W$ will be modeled via the geometry of the map $\pi: \M \rightarrow \B \M$ and the $(S^1, W)$-multi-fiber structure,  and via the metric $\g$ or its Einstein curvature $\w G$. The idea is also that what passes to the quotient can be neglected.

\subsubsection{Electromagnetic potential.}
We suppose that we have the canonical standard orientation on $S^1$. We can then define:
\begin{greybox}
We define $\w Y$ to be the vector field defined at each $x \in \M$ to be tangent to the fiber $S^1_x$ and such that $\g (\w Y,\w Y)=-1$, with the chosen orientation for $S^1$. Again, $\w F=\dd (\w Y^\flat)$. We always suppose from now on that $\w Y$ is a Killing vector field. (Note once again that if $\w Y$ is Killing and of constant norm, it is necessarily geodesic.) The local diffeomorphisms generated by $\w Y$ are therefore isometries.
\end{greybox}
We also add to the definition of $\g$ being compatible with the given $(S^1,W)$-multi-fiber structure on $M$ the following requirement : for any pair of adapted charts $\phi_i$, $\phi_j$ as defined above, $\forall x \in \U_i\cap \U_j$, 
$\phi_i^*(\partial_t)_x$ and $\phi_j^*(\partial_t)_x$ are timelike and in the same time orientation, i.e. $g(\phi_i^*(\partial_t)_x,\phi_j^*(\partial_t)_x)<0$, where $\partial_t$ is the tangent vector to the canonical coordinates $(t,x,y,z)$ on $\Theta_i \subset \mathbb{R}^4$. This condition gives a "classical" time-orientation on every apparent space-time $H_x$, varying differentially with $x$.

\subsection{General Fluids}
Considering now our basic "space-time" model as being the $5+m$-dimensional manifold $\M$, the horizontal space, representing "classical" 4-dimensional space-time, is $H_x := T_x(S_x^1 \times W_x)^{\bot}$. For a fluid, the important object will be once again the endomorphism field $^e  \w G_{ H}=pr_{H} \circ (^e  \w G_{|  H})$, which is essentially the endomorphism field $^e  \w G$, $\g$-associated to the Einstein curvature $\w G$, restricted to the horizontal space $H$.

We start with the most general definition for a matter fluid : 
\begin{greybox}
\begin{definition}
 A domain $\D$ of $\M$ is a fluid domain if, at every $x \in \D$, the endomorphism $^e G_H=pr_H \circ (^e G_{|H})$ has a timelike 1-dimensional eigenspace $E_{-\mu}$ of eigenvalue $-\mu <0$, where we now consider $G$ as being twice the Einstein curvature tensor: 
 \center{\fbox{$G:=2.\w{Ric}_{\g} - \w S_{\g}\, . \g$}}
\end{definition}
\end{greybox}

We now define \emph{naturally} the following objects:
\begin{itemize}
\item $Y$ is the electromagnetic potential, and $F=\dd (Y^\flat)$ is the electromagnetic field.
\item The vector field $X_0$, already seen, such that at every $x\in \D$, $X_0(x)$ is the unique vector of the eigenspace $E_{-\mu}(x)$ in the chosen orientation, and such that $g(X_0,X_0)=-1$, which can be proven easily to be unique. The vector field $X_0$ will be called the \emph{apparent, or visible,} field of the fluid, and the associated flow, the \emph{apparent, or visible,} flow.
\item The smooth function $\mu :\D \rightarrow \mathbb{R}$ defined by $\mu(x)=\mu_x$ where $-\mu_x$ is the eigenvalue associated to the eigenspace $E_{-\mu}$. It will be called the \emph{energy density} of the fluid.
\item The smooth function $e :\D \rightarrow \mathbb{R}$ defined by $e(x)=G_x(X_0(x),Y_x)$. It will be called the electric charge density of the fluid.
\item The vector field $X=X_0+ \frac{e}{\mu}Y$, timelike, is called the vector field of the fluid, and the associated flow, the flow of the fluid.
\item The time-plane $\mathrm{T}_x$, of dimension 2, is the subspace of $T_xM$ generated by $X_0(x)$ and $Y_x$. Because $Y$ is Killing, it can be shown that $[X_0,Y]=0$, therefore the plane field $\mathrm{T}$ is integrable.
\item The time-tube $\mathcal{T}_x$ is then, at each $x\in \D$, the integral submanifold passing through $x$ of the field $\mathrm{T}$. $\mathcal{T}_x$ is a submanifold of dimension 2, totally timelike. These "tubes" can be seen as a generalization of the flow lines of a fluid. They are oriented by the orientation of $X_0$ and $Y$. 
\end{itemize}

With these definitions, for all $x \in M$, the tensor $G$ restricted to $\mathrm{T}_x$ can be written  
$$G_{|\mathrm{T}_x}=\mu X \ts X+ \beta Y\ts Y$$ where $\beta :\D \rightarrow \mathbb{R}$ is a smooth function.

Then, the tensor field $P=G-G_{|\mathrm{T}_x}$ will be called the fluid pressure. It satisfies $P(X_0,X_0)=P(X_0,Y)=P(Y,Y)=P(X,X)=0$. Therefore $G$ can be written :

 \begin{equation*}
\fbox{ $\begin{split}
G &=\mu X \ts X+ \beta Y\ts Y+P\\
       & =\mu X_0 \ts X_0+ e(X_0 \ts Y+Y \ts X_0)+(\beta+\frac{e^2}{\mu}). Y \ts  Y+P
       \end{split}$}
\end{equation*}

\begin{itemize}
\item The \emph{apparent pressure} $P_v$ will be the pressure $P$ restricted to the horizontal space $H_x$. That is: $\forall Z,Z' \in H_x$, $P_v(Z,Z')=P(Z,Z')$, and $\forall Z \in T_x \D$, $\forall Z' \in T_xS^1 \oplus T_xW_x$, $P_v(Z,Z')=0$ and $P_v(Z,X_0)=0$.

\item The \emph{hidden pressure} is $P_h:=P-P_v$.
\end{itemize}
\begin{greybox}
$G$ can now be written : \fbox{$G= \mu X \ts X+ \beta Y\ts Y+P_v+P_h$}
\end{greybox}

In matrix form, with suitable basis for $T_xH$, $T_xS^1_x$ and $T_x W_x$, $G$ can be written: 

\[
G=\left(\begin{array}{cccc} \mu & \begin{array}{ccc}  0&0&0 \end{array} & e 
\\  \begin{array}{c} 0\\0\\0  \end{array}& P_v & \begin{array}{c} a\\b\\c \end{array} & P_h
\\ e &  \begin{array}{ccc} a&b&c \end{array}  & \gamma 
\\  & P_h & & P_h
\end{array}\right)
\]
where $a,b,c$ are the component of $P_h(Y)$ on $H'  := (T_x(S_x^1 \times W_x) \oplus <X_{0_x}>)^{\bot}$.

The fluid will be called perfect if $P_h(Y)=0$. $G$ can then be written: 

\[
G=\left(\begin{array}{cccc} \mu & \begin{array}{ccc}  0&0&0 \end{array} & e 
\\  \begin{array}{c} 0\\0\\0  \end{array}& P_v & \begin{array}{c} 0\\0\\0 \end{array} & P_h
\\ e &  \begin{array}{ccc} 0&0&0 \end{array}  & \gamma & 0
\\  & P_h & 0 & P_h
\end{array}\right)
\]

This will now be our general model for a fluid.
One can then obtain the following theorem, applying the same scheme as the one shown in 5 dimensions, using Bianchi identity, and computing $\Div F$.  Remember that we identify the 1-form $\Div \w T$ and $(\Div \w T)^\sharp$ and that we note $\la V,W \ra:=\g (V,W)$.

\begin{greybox}
\begin{theorem} \emph{Equations for the spacetime dynamics of fluids.}

If $\D$ is a fluid domain as above, Bianchi identity gives : 
\begin{itemize}
\item Energy Conservation Laws : $\Div (\mu X)=\Div (\mu X_0)=\la X_0,\Div P \ra$ 
\item Electric Charge Conservation Law : $\Div (e X)=\Div (eX_0)=\Div (^eP(Y))= \la Y,\Div P \ra$
\item Furthermore : $\mu^2 X(\frac{e}{\mu })=\mu .\la Y, \Div P\ra -e.\la X_0,\Div P \ra$ and $\la Y,\Div P\ra =\Div (\, ^eP(Y))$

\item Motion Equations : 
\begin{itemize}
\item For the fluid : $\mu \wnab_X X=-\Div P-\la X_0,\Div P \ra X$
\item For the apparent fluid : $\mu \wnab_{X_0}X_0 =e.\, ^eF(X_0)-pr_{\mathrm{T}^{\bot}}(\Div P)$
\end{itemize}
\item Maxwell Equations : $\dd F=0$, and $\Div F=e.X_0+1/2|F|_g .Y-\, ^eP(Y)$
\end{itemize}
\end{theorem}
\end{greybox}

\begin{greybox}
The theorem we gave in 5 dimension can be seen as a special case where $P_h=0$. For dust, $P_v=0$. 
In this last theorem, the pressure $P$ can be split everywhere into $P=P_v + P_h$, to show physical effects due to the three "classical" dimensions, $P_v$, and those due to the extra "hidden or small" dimensions, $P_h$. This theorem therefore shows that $P_h$, the hidden pressure, is a possible way  to model or encode deviations from standard 4-dimensional General Relativity, or "dark" effects such as dark matter or energy.
\end{greybox}

Once again, this theorem is a purely geometrical fact, based on Bianchi identity. This proof is just a technical generalization of the proof made in the case of dust in 5 dimensions. The essential ideas are the same : 1/ computing $\Div G$ and projecting it on $X_0$, $Y$ and $H$ for conservation and motion equations, 2/ computing $\Div F$ for the second Maxwell equation, and 3/ computing $\wnab_X X$. The technicalities are due to the computations linked to the presence of the pressure $P$.  \\
\emph{In particular, if we suppose that $P_h=0$, we recover the classical equations of general relativity with electromagnetism for a perfect fluid.}

\subsection{Special Fluids}
The definitions we are now going to give are here to obtain the classical physics equations of general relativity. As these equation will be given in 5+m dimensions, it is by projection on the apparent space-time $H_x$, naturally isometric to $\T_{\pi(x)}\B\M$, than the comparison will have to be made.  In particular, if $W=\{p\}$, or $m=0$, we recover the results of section 4 concerning 5-dimensional spacetime.

\begin{greybox}
\begin{definition}Special fluids :
\begin{itemize}
\item A fluid domain $\D$ is a \emph{perfect fluid} domain, if $^eP(Y)=0$. Note that $^eP(Y)=\, ^eP_h(Y)$.
\item A fluid domain $\D$ is a \emph{perfect isentropic fluid} domain, if it is a perfect fluid domain such that at each point $x \in \D$, the pressure tensor $P_x$ is proportional to $g_x-(g_x)_{|\mathrm{T}_x}$, that is $P=p(g+X_0 \ts X_0+Y \ts Y)$ for some smooth function $p:\D \rightarrow \mathbb{R}$.
\item A fluid domain $\D$ is a \emph{pluperfect fluid} domain, possibly electrically charged, if it is a perfect fluid such that $\Div P=0$. For example, dust, for which $P=0$.
\item A pluperfect fluid without electromagnetism, is a pluperfect fluid such that $e=0$ and $F=0$.
\end{itemize}
\end{definition}
\end{greybox}

Applying the previous theorem for general fluids to these special fluids gives:
\begin{greybox}
\begin{theorem}
 For a \emph{Perfect Fluid}, the following equations are valid:
\begin{itemize}
\item Energy Conservation Laws : $\Div (\mu X)=\Div (\mu X_0)=\la X_0,\Div P \ra$. 
\item Electric Charge Conservation Law : $\Div (e X)=\Div (eX_0)=0$
\item Furthermore : $\mu^2 X(\frac{e}{\mu })=\mu^2 X_0(\frac{e}{\mu })=-e.\la X_0,\Div P \ra$
\item Motion Equations : 
\begin{itemize}
\item For the fluid : $\mu \wnab_X X=-\Div P- \la X_0,\Div P\ra X$
\item For the apparent fluid : $\mu \wnab_{X_0}X_0 =e.\, ^eF(X_0)-\Div P- \la X_0,\Div P \ra X_0$
\end{itemize}
\item Maxwell Equations : $\dd F=0$, and $\Div F=e.X_0+1/2|F|_g Y$.\\
 In particular, $pr_{H_x}(\Div F)=eX_0$
\end{itemize}
\end{theorem}
\end{greybox}

\begin{greybox}
\begin{theorem}
 For an \emph{Isentropic Perfect Fluid} :
\begin{itemize}
\item Energy Conservation Laws : $\Div \mu X_0+p.\Div X_0=0$
\item Electric Charge Conservation Law : $\Div (e X)=\Div (eX_0)=0$
\item Furthermore : $\mu^2 X(\frac{e}{\mu })=\mu^2 X_0(\frac{e}{\mu })=e.p.\Div X_0$
\item Motion Equations : $(\mu +p).\wnab_{X_0}X_0=e.^eF(X_0)-grad _g (p) - X_0(p).X_0$

\item Maxwell Equations : $\dd F=0$, and $\Div F=e.X_0+1/2|F|_g Y$.\\
 In particular, $pr_{H_x}(\Div F)=eX_0$.
\end{itemize}
These are exactly the equations of general relativity for isentropic charged fluids.
\end{theorem}
\end{greybox}

\begin{greybox}
\begin{theorem}
For a \emph{Pluperfect Fluid (e.g. electrically charged dust)} :
\begin{itemize}
\item Energy Conservation Laws : $\Div (\mu X)=\Div (\mu X_0)=0$. 
\item Electric Charge Conservation Law : $\Div (e X)=\Div (eX_0)=0$.
\item Furthermore : $X(\frac{e}{\mu })=X_0(\frac{e}{\mu })=0$.
\item Motion Equations : 
\begin{itemize}
\item For the fluid : $\wnab_X X=0$
\item For the apparent fluid : $\mu \wnab_{X_0}X_0 =e.\, ^eF(X_0)$
\end{itemize}
That is, $X$ is a geodesic vector field, even if the fluid has an electrical charge. Of course, if $e=0$, $X_0$ is geodesic.
\item Maxwell Equations : $\dd F=0$, and $\Div F=e.X_0+1/2|F|_g .Y$
\end{itemize}
\end{theorem}
\end{greybox}
As we already said, comparing this last theorem considering dust with the one for general fluids, we see how the extra dimensions, (those of $W$, above 5), and the hidden pressure, $P_h$, give deviations from our 5-dimensional model of general relativity with electromagnetism.
\\

In the motion equations, $\wnab_{X_0}X_0$ and $^eF(X_0)$ are $\g$-orthogonal to $Y$, but not necessarily to $W$. If one wants to guaranty that $\wnab_{X_0}X_0$ and $^eF(X_0)$ belong to $H_x$, one can impose the following requirement :

\emph{The submanifolds $W_x$ are parallel along the geodesic circles $S^1_x$ : $\forall x\in \D$, $\forall Z \in \T_xW_x$, $\forall x'\in S^1_x$, the parallel transport of $Z$ along $S^1_x$ is tangent to $W_{x'}$.}

It is then quickly verified that under this condition, $\forall x\in \D$, $^eF(X_0) \in H_x$.

\section{Newtonian and electromagnetic potential in 5+m dimensions.}
We now change our point of view, abandoning fluids, and defining region of spacetime empty of matter but considered as potentials, and testing their properties by establishing the geodesic motion of test particles. All what follows in this chapter is due to Michel Vaugon.
\begin{greybox}
\begin{definition}
A domain $\D$ of $\M$ is a \emph{potential domain} if $\forall x \in \D$, $G_{H_x}=0$ and $pr_{H_x}\,^eG(Y)=0$.
\end{definition}
\end{greybox}
These domain are therefore extensions of fluid domain in which the energy density, the charge density and the apparent pressure $P_v$ are null. Thus, there is no canonically defined vector fields $X$ or $X_0$. Only remains the electromagnetic objects $Y$ and $F=dY^*$. The tensor $G$ is then now equal to the hidden pressure, which therefore satisfies $\Div P_h=0$.

The above theorems give:
\begin{greybox}
\begin{theorem}
In a \emph{Potential domain} we have the Second Maxwell equation:
$$\Div F= 1/2 |F|_g.Y-\, ^eG(Y)$$
In particular, as $pr_H\, ^eG(Y)=0$, we have : $pr_H \Div F=0$.

The first Maxwell equation $\dd F=0$ is obvious as $F=dY^\flat$.
\end{theorem}
\end{greybox}
Potential domains are very important as the knowledge of their geodesics gives the motion curves of "test particles" placed in these potentials, when it is considered that their effect on the geometry of the domain can be neglected. Indeed, if one "introduce" a test particle in a potential domain, it can then be considered as a dust fluid where the energy density $\mu $ is not zero, this "spatial" domain being very limited. If we consider that outside this domain, the geometry of the potential domain is not affected, the flow field $X$ of the fluid, defined when $\mu \neq 0$, is a geodesic field according to theorem 11. The curves of the flow $X$ can thus be considered approximately as the geodesics of the potential domain. Furthermore, the quotient $\frac{e}{\mu }$ can be considered as the quotient of the charge by the mass of the test particle. This is a classical method in general relativity for test particles without electrical charge, as for example in the Schwarzschild solution. The remarkable thing in our setting is that this principle now applies even for test objects with an electrical charge, but in a "space-time" of dimension greater or equal to 5. In this case, the apparent trajectory is determined from the apparent field $X_0$, itself being obtained from the geodesic field $X$. Of course, $X=X_0$ when the charge is zero.

We shall now present examples of such domains, with the big advantage of being given with the exact metric tensor $g$. Precise computations of geodesics will be given, obtained with the help of Mapple software, or, better as it if free, SAGE software.

\subsection{Preliminary geometrical setting.}
The circle $S^1(\delta)$ is defined as $S^1(\delta)=\mathbb{R}/2\pi \delta \mathbb{Z}$. The natural surjection $ \Pi : \mathbb{R} \rightarrow 2\pi \delta \mathbb{Z}$ gives a natural origin $P=\Pi (0)$, a natural orientation (that of $\mathbb{R}$ carried by $\Pi$), a natural coordinate $u \in ]0, 2\pi[$ for $\T u =\Pi(u) \in S^1(\delta)-\{P\}$, and a metric $g_{S^1(\delta)}$, the metric of $\mathbb{R}$ quotiented by $\Pi$.

We define the torus $T^n(r_1,...,r_n)=S^1(r_1)\times ...\times S^1(r_n)$, which therefore carries by using the above definition for $S^1(\delta)$, a natural origin, a natural coordinate system, and a natural metric, the product metric : $g_{T^n(r_1,...,r_n)}=g_{S^1(r_1)} \times ... \times g_{S^1(r_n)} $.

We shall call these definitions the "standard setting" on $S^1$ or $T^n$.

 Let $\Theta$ be a open set in $\mathbb{R}^4$ and $$\mathcal{C}=\Theta \times S^1(\delta ) \times  T^{n-5}(r_1,...,r_{n-5})$$
$\mathcal{C}$ will be called a standard cell. The standard coordinate on $\mathcal{C}$ will be denoted by $(t,x,y,z,u,v_1,...,v_{n-5})$, where $(t,x,y,z) \in \Theta$. The standard metric on $\mathcal{C}$ is the product metric $g_0 := g_{\Theta} \times (-g_{S^1(\delta)}) \times g_{T^{n-5}(r_1,...,r_{n-5})}$, where $g_{\Theta}$ is the Minkowski metric on $\Theta \subset \mathbb{R}^4$. In standard coordinates, it is written $$g_0=-dt^2+dx^2+dy^2+dz^2-du^2+dv_1^2+...+dv_{n-5}^2$$ This metric will be called the Minkowski metric of the standard cell $\mathcal{C}$. Its signature is everywhere $(-,+,+,+,-,+,...+)$.

We now consider on a standard cell $\mathcal{C}$, the following two objects : 

-A function $V : \mathcal{C} \rightarrow \mathbb{R}$ where $V$ is a function of the variables $(x,y,z)$.

-A 1-form $\gamma := \phi dt +A_1dx+A_2dy+A_3dz$ on $\mathcal{C}$, where the functions $\phi , A_1,A_2,A_3$ are functions of $(t,x,y,z)$.

We will denote by $F$ the 2-form $F:= d\gamma$ on $\mathcal{C}$.

We shall use the classical terminology : $V$ will be called the Newtonian potential, $\gamma$ the electromagnetic potential ($\phi$ the electric potential, $(A_1,A_2,A_3)$ the magnetic potential), and $F$ the electromagnetic field 2-form.

We now define the pseudo-Riemannian tensor $g_1$ by 
$$g_1:=g_0-2V. \mathcal{N}_1 \ts \mathcal{N}_1+(\gamma \ts \mathcal{N}_2 +\mathcal{N}_2 \ts \gamma)$$
where $\mathcal{N}_1:=dt+dv_1$, and $\mathcal{N}_2:=du+dv_2$.
In standard coordinates, the matrix of $g_1$ is therefore (considering $n=8$) : 
\[
g_1=\left(\begin{array}{cccccccc}
-1-2V & 0 & 0 & 0 & \phi &-2V& \phi & 0 \\
0 & 1 & 0 & 0 & A_1 & 0 & A_1 & 0 \\
0 & 0 & 1 & 0 & A_2 & 0 & A_2 & 0 \\
0 & 0 & 0 & 1 & A_3 & 0 & A_3 & 0 \\
\phi  & A_1 & A_2 & A_3 & -1 & 0 & 0 & 0 \\
-2V & 0 & 0 & 0 & 0 & -2V+1 & 0 & 0 \\
\phi  &A_1 & A_2 & A_3 & 0 & 0 & 1 &0 \\
0 & 0 & 0 & 0 & 0 & 0 & 0 & 1
\end{array}\right)
\]

We also consider the following two particular cases:

$g_N:=g_0-2V. \mathcal{N}_1 \ts \mathcal{N}_1$, that is $g_1$ where $\gamma=0$,

and

$g_E:=g_0+(\gamma \ts \mathcal{N}_2 +\mathcal{N}_2 \ts \gamma)$, that is $g_1$ where $V=0$.
\\

The following results can then be shown :

\begin{itemize}
\item $det(g_1)=det(g_0)=1$
\item the 1-forms $\mathcal{N}_1$ and $\mathcal{N}_2$ are isotropic both for $g_0$ and $g_1$.
\item The Laplacians $\Delta_{g_0}V=\Delta_{g_1}V$, and will therefore be denoted by $\Delta V$.
\item $\w{Ric}_{g_N}=(\Delta V)\mathcal{N}_1 \ts \mathcal{N}_1$.
\item $\w{Ric}_{g_E}=1/2(H.\mathcal{N}_2 \ts \mathcal{N}_2-(\nabla_{g_0}\cdot F) \ts \mathcal{N}_2+\mathcal{N}_2 \ts (\nabla_{g_0}\cdot F) )$
\item The Scalar curvatures : $S_{g_N}=S_{g_E}=0$ 
\item $G_N=2(\Delta V)\mathcal{N}_1 \ts \mathcal{N}_1$
\item $G_E=H.\mathcal{N}_2 \ts \mathcal{N}_2-(\nabla_{g_0}\cdot F) \ts \mathcal{N}_2+\mathcal{N}_2 \ts (\nabla_{g_0}\cdot F)$.
\item For any $(P,Q) \in \Theta \times T^{n-5}$, the circle $\{P\}\times S^1(\delta)\times \{Q\}$ is a timelike geodesic both for $g_0$ and $g_1$.
\end{itemize}
Here $\nabla_{g_0}\cdot F=\partial_i(g_0^{ik}F_{kj})$ and  $H:=|F|_{g_0}=g_0^{ik}g_0^{jl}F_{kl}F_{ij}$.

We then consider the vector field $Y$ tangent to these circles, oriented by the standard orientation, and such that $g_1(Y,Y)=-1$, (in fact $Y=\partial_u$). We have the following results:
\begin{itemize}
\item The electromagnetic potential $\gamma$ is the 1-form $g_E$-associated to $Y$, i.e. $\gamma_i = (g_E)_{ij}Y^j$. 
\item $Y$ is a Killing vector field for $g_E$, and $\wnab Y=0$ for $g_N$.
\end{itemize}

These results are not difficult, but obviously requires some tedious computations. A mathematical software like Sage is a big help...
\\

We now make the following important remark : Both $g_N$ and $g_E$ are of the form $g=g_0+h$, where of course $h$ is a bilinear form. If we consider the endomorphisms field $^e h$ associated to $h$ by $g_0$, it was proved by Michel Vaugon that in both cases, $^e h$ is nilpotent:
\begin{greybox}
We say that an endomorphisms field $^e h$ is \defin{nilpotent} of index $p\in \mathbb{N}$ if $\forall x\in \M$ and $\forall q\geq p$, the endomorphism $^e h_x$ of $\T_x \mathcal{C}$ satisfies $(^e h_x)^q =0$, and if there exists $x \in \mathcal{C}$ such that $(^e h_x)^{p-1}\not =0$.
\end{greybox}
 \begin{greybox}
 Let us write $g_N=g_0+h_N$ and $g_E=g_0+h_E$. Then $^e h_N$ is nilpotent of index 2, and $^e h_E$ is nilpotent of index 3.
 \end{greybox}
It is of course essential that the metric $g_0$ is not positive definite, otherwise, a nilpotent endomorphism can only be null.

This lead Michel Vaugon to define \defin{active potential} as potential domains where the metric can be written in the form $g=g_0+h$, with $^e h$ nilpotent of index $p\geq 2$. See [ref]. Note also that he defined there domains with more general fibers than $S^1 \times$ torus. The definition of potential domains can therefore be extended to multi-fibers bundle of the form $S^1 \times W$ with $W$ compact.

\subsection{Newtonian Potential}
We consider in this subsection a manifold $(\mathcal{D},g)$ equipped with a $(S^1(\delta ) \times  T^{n-5}(r_1,...,r_{n-5}))$-multi-fiber structure where $(\mathcal{D},g)$ is isometric, via some $\varphi :\mathcal{D} \rightarrow \mathcal{C}$, to the standard cell $(\mathcal{C},g_N)$ defined in the previous section. From now on, we identify $(\mathcal{D},g)$ with $(\mathcal{C},g_N)$.

Considering $(\mathcal{D},g)$ as a potential domain means that we suppose that $\forall x \in \mathcal{D}$, $G_{Nx|H_x}=0$. From the results of the preliminary setting, this means that $\Delta V=0$, that is, $G_N=0$.

We now study the geodesics of this domain.

In the standard coordinate system $(t,x,y,z,u,v_1,...,v_{n-5})$, the Christoffel symbols of $g_N$ are :
$$\Gamma_{1j}^1=\Gamma_{6j}^1=-\Gamma_{1j}^6=-\Gamma_{6j}^6=\Gamma_{11}^j=\Gamma_{16}^j=\Gamma_{66}^j=\partial_j V$$
where $\partial_j:=\partial/\partial x_j$, and $x_j$ is the $j-th$ coordinate in $(t,x,y,z,u,v_1,...,v_{n-5})$. The other symbols are all zero, except of course the "symmetric" $\Gamma_{ij}^k=\Gamma_{ji}^k$.

Let $\alpha (s)=(t(s),x(s),y(s),z(s),u(s),v_1(s),...,v_{n-5}(s))$ be a geodesic of $(\mathcal{D},g)$, parametrized by $s \in \mathbb{R}$.

Using the above Christoffel symbols, it satisfies the following equations :
$$t''+2(x'.\partial_x V+y'.\partial_yV+z'.\partial_z V)(t'+v_1')=0$$
$$v_1''-2(x'.\partial_x V+y'.\partial_yV+z'.\partial_z V)(t'+v_1')=0$$
$$x''+(t'+v_1')^2.(\partial_x V)=0$$
$$y''+(t'+v_1')^2.(\partial_y V)=0$$
$$z''+(t'+v_1')^2.(\partial_z V)=0$$
$$u''=v_2''=...=v_{n-5}''=0$$
where $t'=t'(s)$, $\partial_x V=(\partial_x V)_{\alpha(s)}$, etc...

The first two equations can be written :
$$t''+2(V(\alpha(s)))'.(t'+v_1')=0$$
$$v_1''-2(V(\alpha(s)))'.(t'+v_1')=0$$
Thus, $t''+v_1''=0$, and $t'+v_1'=k$. Up to a reparametrization, we can suppose $k=1$. ($k=0$ is not interesting for us). Then :
$$t''+2(V(\alpha(s)))'=0$$
$$v_1''-2(V(\alpha(s)))'=0$$
and 
$$t'=-2V(\alpha(s))+c$$
$$v_1'=2V(\alpha(s))+1-c.$$
The three next equations give : $x''=-\partial_x V$, $y''=-\partial_y V$, $z''=-\partial_z V$, that is :
$$(x(s),y(s),z(s))''=-(\nabla V)_{\alpha(s)}$$
which is exactly Poisson equation in classical physics when $V$ is a Newtonian potential ($\Delta V=0$) and when $(x(s),y(s),z(s))$ represents the trajectory of a test particle of mass $m$ in such a potential, but here \emph{considering that $s$ is the time parameter.}

However, if we suppose that $V=o(1)$, and if we consider only the geodesics for which $v_1'(s)=o(1)$, that is those for which the speed component along the circle $S^1(r_1)$ is small compared to 1, (the speed of light), we see from the above equations that $t'(s)=1+o(1)$. In this case, this means that the parameter $s$ is very close to the "time" $t$, and that the Poisson equation is almost classically satisfied.

Remarks concerning this "Newtonian potential":

Let us set, in the standard coordinates, $V=-\frac{m}{r}$, where $r=\sqrt{x^2+y^2+z^2}$ and $m$ is a positive constant. The standard cell is then "space-symmetric" for the usual coordinates $(x,y,z)$. We have $\Delta V=0$, and we just saw that for the metric $g_N=g_0+2\frac{m}{r} \mathcal{N}_1 \ts \mathcal{N}_1$, the geodesics $\alpha (s)=(t(s),x(s),y(s),z(s),u(s),v_1(s),...,v_{n-5}(s))$ satisfies (at least those of interest): 
$$(x(s),y(s),z(s))''=-(\nabla V)_{\alpha (s)}=\frac{m}{r(\alpha (s))^2}. $$
We therefore conclude, as in classical mechanics, that the image of these geodesics projected on classical space $(x,y,z)$ are exacly conics for which $(0,0,0)$ is a focal point, and for which Kepler laws are valid, but when considering the parameter $s$ instead of "time" $t$ of the coordinate system, which can differ greatly from $s$ if $V \neq o(1)$, that is if $r$ is close to $0$.

Because $g_N (\partial_t, \partial_t)=-1+2m/r$, the vector field $\partial_t$ is timelike if $r>2m$, null if $r=2m$, and spacelike if $0<r<2m$.
The critical radius $r=2m$ corresponds to the Schwarschild radius, and therefore suggests to compare this Newtonian potential domain to the classical Schwarschild domain which could be defined here to be a standard cell $(\mathcal{C},g_S)$ :
$$\mathcal{C}=\Theta \times S^1(\delta ) \times  T^{n-5}(r_1,...,r_{n-5})$$
where $\Theta =\mathbb{R}\times \mathbb{R}^{3*}$, and $g_S$ is the product metric $g_S=g_{\Theta} \times g_W$, with $g_W$ the standard metric on $W= S^1(\delta ) \times  T^{n-5}(r_1,...,r_{n-5})$, and $g_{\Theta}$ is the classical Schwarschild metric, written in spherical coordinate $(t,r,\varphi, \phi)$ on $\mathbb{R}\times ]2m, + \infty [ \times S^2 \backsim \mathbb{R}\times \mathbb{R}^{3*}$ :
$$g_{\Theta}(t,r,\varphi, \phi)=(-1+2m/r)dt^2+(1-2m/r)^{-1}dr^2+r^2(d\varphi^2+ sin^2\varphi d\phi^2).$$

The purpose of the product $g_S=g_{\Theta} \times g_W$ is only to carry the classical 4-dimensional Schwarschild metric into our 5+m-dimensional setting. We could also consider the Schwarschild domain extended to $0<r\leqslant 2m$.

Let us compare some properties of the domains $(\mathcal{C},g_N)$ and $(\mathcal{C},g_S)$.
\begin{itemize}
\item Both Ricci curvatures : $\w{Ric}_{g_N}$ and $\w{Ric}_{g_S}$ are zero.
\item For $r>>2m$, the geodesics of $(\mathcal{C},g_N)$ and the timelike geodesics of $(\mathcal{C},g_S)$ (whose components are constant on $S^1(\delta ) \times  T^{n-5}(r_1,...,r_{n-5})$) give, with very good approximation, the trajectories of test particles around a body of mass $m$ with spherical symmetry in $(x,y,z)$ space, computed in classical Newtonian mechanics.
\end{itemize}

One can also notice that the coefficient $(-1+2m/r)$ in front of $dt^2$ in $g_N$ is the same as that of $g_S$. However, for $g_N$ the potential $2m/r$ is perturbating the "small" dimensions of $T^{n-5}$ without affecting the classical dimensions $(x,y,z)$, whereas for $g_S$, the potential $2m/r$ perturbates the $(x,y,z)$ dimensions, without affecting the "small" dimensions.

The two cells $(\mathcal{C},g_N)$ and $(\mathcal{C},g_S)$ could therefore be considered as extreme particular cases of a family of domains resembling Newtonian potentials, for which the potential is affecting every dimensions, and which still satisfy the two above conditions.

\begin{greybox}
Just as for general fluids in a $(5+k)$-dimensional spacetime of section 5, this Newtonian potential shows that  the extra "hidden or small" dimensions are a way  to model or encode deviations from standard 4-dimensional General Relativity, or "dark" effects such as dark matter or energy.
\end{greybox}

\subsection{Electromagnetic Potential}
We now consider a a manifold $(\mathcal{D},g)$ equipped with a $(S^1(\delta ) \times  T^{n-5}(r_1,...,r_{n-5}))$-multi-fiber structure  where $(\mathcal{D},g)$ is isometric, via some $\varphi :\mathcal{D} \rightarrow \mathcal{C}$, to the standard cell $(\mathcal{C},g_E)$ defined in section 16.6.1. From now on, we identify $(\mathcal{D},g)$ with $(\mathcal{C},g_E)$.

By the given definition of a potential domain, and the results of section 16.6.1,\\
 $\nabla_{g_0} \cdot F=0$. Thus : 
$$G_E=H. \mathcal{N}_2 \ts \mathcal{N}_2.$$

 Let us study the geodesics of this domain.

In the standard coordinate system $(t,x,y,z,u,v_1,...,v_{n-5})$, to simplify the numeration of the Christoffel symbols, we set $\mathcal{N}_2=du+dv_1$, (instead of $du+dv_2$). In this coordinate system, the Christoffel symbols of $g_E$ satisfy :
\begin{itemize}
\item $\forall k$, $\Gamma_{ij}^k=0$ if $i,j \geqslant 5$, $i$ or $j$ $>6$.
\item $\forall i,j$, $\Gamma_{ij}^k=0$ if $k>6$.
\item $\Gamma_{ij}^k=0$ if $i,j,k<5$.
\item  $\Gamma_{i5}^k=\Gamma_{i6}^k=1/2g^{kk}(\partial_i g_{5k}-\partial_k g_{i5})=1/2g^{kk}(\partial_i g_{6k}-\partial_k g_{i6})$ if $k<5$.
\item $\forall i,j$, $\Gamma_{ij}^5+\Gamma_{ij}^6=0$.
\end{itemize}

Let $\alpha (s)=(t(s),x(s),y(s),z(s),u(s),v_1(s),...,v_{n-5}(s))$ be a geodesic of $(\mathcal{D},g)$, parametrized by $s \in \mathbb{R}$.
Using the above Christoffel symbols, it satisfies the following equations :

If $k<5$, (i) : $x_k'' (s)+2(\sum_{i=1}^4 \Gamma_{i5}^k x_i' (s))(x_5'(s)+x_6' (s))=0$, where $x_j$ is the $j-th$ coordinate in $(t,x,y,z,u,v_1,...,v_{n-5})$, and $\Gamma_{i5}^k=\Gamma_{i5}^k (\alpha(s))$.

Furthermore, $u''+\sum_{i,j}\Gamma_{ij}^5 x_i'x_j'=0$, and $v_1''+\sum_{i,j}\Gamma_{ij}^6 x_i'x_j'=0$.

From this we get : $(u+v_1)''(s)=0$, and $u'+v_1'=c$. (i) can therefore be rewritten : $x_k''=-2c(\sum_{i=1}^4 \Gamma_{i5}^k x_i' (s))$.

But $Y$ is a Killing vector field, so $F^k_{\, \, \, i}=-2\nabla_i Y^k=-2(\partial_i Y^k +\Gamma_{il}^k Y^l)=-2\Gamma_{i5}^k$. So : 
$$x_k''(s)= c \sum_{i=1}^4 F^k _{\, \, \, i}\,\, x_i'(s)$$
as $F^k \, _i=0$ if $i\geqslant 5$.

Now, $\alpha(s)$ can be parametrized so that $g(\dot \alpha(s),\dot \alpha(s))=-1$, and $\dot \alpha(s)_{|H_x}$ is in the time orientation given by $t$. We will of course call $s$ the proper time of the geodesic.

Denoting by $X(s)=(t'(s),x'(s),y'(s),z'(s))$ the vector corresponding to the first 4 components of $\dot \alpha(s)$, the above equations can be written : 
$$X'(s)=c.\, ^eF(X(s)).$$

We recover the classical equation of the motion of a particle of mass $m$ and of charge $q$ in an electromagnetic field $F$ when we set $c=q/m$ and when $s$ indeed represents the proper time of the particle.

Here, $c=q/m=u'+v_1'$ is a characteristic data of the geodesic on the "small" dimensions.

If the speeds $|x_k '(s)|=o(1)$ for $k \neq 1,5,6$ and if $|u'|=|v_1'|+o(1)$, then $t'=1+o(1)$ and the "time" $t$ of the coordinate system is very close to the proper time of the geodesics; this corresponds to the classical approximation on non relativistic classical mechanics.


\end{document}